\documentclass[twocolumn,3p]{elsarticle}
\usepackage{subfigure}

\usepackage{natbib}
\usepackage{captcont}
\usepackage[final,		% override "draft" which means "do nothing"
	colorlinks,			% rather than outlining them in boxes
	linktocpage,		% link just the pages, not the full titles
	linkcolor=blue,		% override truly awful color choices
	citecolor=blue,		% (ditto)
	urlcolor=blue,		% (ditto)
	breaklinks=true,	% so that links wrap around to multiple lines if necessary
	]{hyperref}
\usepackage[all]{hypcap}	% to get figure links correct

\begin{document}

%%%%%%%%%%%%%%%%%%%%%%%%%%%%%%%%%%%%%%%%%%%%%%%%%%%%%%%%%%%%

\begin{frontmatter}

\title{TARGET: A multi-channel digitizer chip for very-high-energy gamma-ray telescopes}
%\title{TARGET: A multi-channel digitizer chip for imaging atmospheric Cherenkov telescopes}
%\title{Design and Performance of the TeV Array Readout with GSa/s sampling and Event Trigger (TARGET) ASIC}
%%% author list %%%%%%%%%%%%%%%%%%%%%%%%%%%%%%%%%%%%%%%%%%%%
\author[KAVLI]{K.~Bechtol\corref{corauths}}
\author[KAVLI]{S.~Funk}
\author[ISAS,KAVLI]{A.~Okumura\corref{corauths}}
\author[UH]{L. L.~Ruckman}
\author[KAVLI]{A.~Simons}
\author[KAVLI]{H.~Tajima}
\author[KAVLI]{J.~Vandenbroucke\corref{corauths}}
\author[UH]{G. S.~Varner}

% Old corauth section
%\corauth[bechtol]{Corresponding author. Tel.: 650-926-4707.}
%\ead{bechtol@slac.stanford.edu}
%\corauth[vandenbroucke]{Corresponding author. Tel.: 650-926-4760.}
%\ead{justinv@slac.stanford.edu}
%\corauth[okumura]{Corresponding author. Tel.: +81 50-3362-8510.}
%\ead{oxon@astro.isas.jaxa.jp}

%\corauth[bechtol]{Corresponding author. bechtol@slac.stanford.edu}
%\ead{bechtol@slac.stanford.edu}
%\corauth[okumura]{Corresponding author. oxon@astro.isas.jaxa.jp}
%\ead{oxon@astro.isas.jaxa.jp}
%\corauth[vandenbroucke]{Corresponding author. justinv@stanford.edu}
%\ead{justinv@slac.stanford.edu}
\cortext[corauths]{Corresponding authors: bechtol@slac.stanford.edu, oxon@astro.isas.jaxa.jp, justinv@stanford.edu.}

\address[KAVLI]{W. W. Hansen Experimental Physics Laboratory, Kavli Institute for Particle Astrophysics and Cosmology, Department of Physics and SLAC National Accelerator Laboratory, Stanford University, Stanford, CA 94305, USA}

\address[UH]{Department of Physics and Astronomy, University of Hawaii, 2505 Correa Road, Honolulu HI 96822, USA}

\address[ISAS]{Institute of Space and Astronautical Science, JAXA, 3-1-1 Yoshinodai, Chuo-ku, Sagamihara, Kanagawa 252-5210, Japan}

%%%%%%%%%%%%%%%%%%%%%%%%%%% Abstract %%%%%%%%%%%%%%%%%%%%%%%%%%%%%%%%%%%%%%%%

\begin{abstract}
The next-generation very-high-energy (VHE) gamma-ray observatory, the Cherenkov Telescope Array, will feature dozens of imaging atmospheric Cherenkov telescopes (IACTs), each with thousands of pixels of photo-sensors.  To be affordable and reliable, reading out such an array of up to a million channels requires event recording technology that is highly integrated and modular, with a low cost per channel.  We present the design and performance of a chip targeted to this application: the TeV Array Readout with GSa/s sampling and Event Trigger (TARGET).  This application-specific integrated circuit (ASIC) has 16 parallel input channels, a 4096-sample buffer for each channel, adjustable input termination, self-trigger functionality, and tight window-selected readout.  We report the performance of the first-generation version of this chip (TARGET~1) in terms of sampling frequency, power consumption, dynamic range, current-mode gain, analog bandwidth, and cross talk.  The large number of channels per chip allows a low cost per channel ($\$10$ to $\$20$ including front-end and back-end electronics but not including photosensors) to be achieved with a TARGET-based IACT readout system.  In addition to basic performance parameters of the TARGET~1 chip, we present a camera module prototype as well as a second-generation chip (TARGET~2), both of which have now been produced.
%Although TARGET was designed to meet the needs of ground-based very-high-energy gamma-ray astronomy, it could also be used for a number of other applications.
\end{abstract}

%%%%%%%%%%%%%%%%%%%%%% keywords and PACS %%%%%%%%%%%%%%%%%%%%%%%%%%%%%%%%%%%%

\end{frontmatter}

% For PACS codes see
% http://www.aip.org/pacs/pacs2010/individuals/pacs2010_regular_edition/alpha_index.html
% 29.40.Ka	Cherenkov detectors					kept it
% 29.40..Mc	Scintillation detectors					removed it
% 87.57.uk	Positron Emission Tomography			removed it
% 95.55.Ka	X- and ?-ray telescopes and instrumentation	added it

%%%%%%%%%%%%%%%%%%%%%% Introduction  %%%%%%%%%%%%%%%%%%%%%%%%%%%%%%%%%%%%

\section{Introduction}
\label{sec:introduction}

Next-generation imaging atmospheric Cherenkov telescope (IACT) instruments such as the Cherenkov Telescope Array (CTA~\cite{CTA}) aim to achieve a 10-fold increase in sensitivity over the current generation of Cherenkov telescopes for $\gamma$-ray observations in the energy band from 40 GeV to 200 TeV.  Reaching this level of performance will require 50-100 telescopes with as many as one million total electronics channels to read out.  Specifically, short-duration ($\sim$10 ns) output pulses from an array of photodetectors at the focal plane of each telescope must be digitized and read out by the camera front-end electronics at sustained rates up to 10 kHz\footnote{Typically all channels of a telescope are read out together, so this is the rate of both single-telescope and single-pixel readout.}.  Trigger rates of individual pixels can be much higher and the decision of whether to digitize and read out a telescope is made by analyzing trigger signals within each telescope (``camera trigger''~\cite{magicTrigger,magicSumTrigger,veritasTrigger}) and among multiple telescopes (``array trigger''~\cite{hessTrigger}).

Producing and operating the readout electronics for an array with so many channels will require individual components with low cost and high reliability compared to the current generation of telescopes.  Several design concepts are under consideration to reduce the cost of the camera electronics while improving their performance.  Systems based on multi-channel waveform sampling ASICs are leading candidates for the front end of such a low-cost electronics system.  We developed the TeV Array Readout with GSa/s sampling and Event Trigger (TARGET) application-specific integrated circuit (ASIC) prototype, optimized specifically for the CTA application.  The TARGET chip can both form its own triggers and receive external triggers from a companion field-programmable gate array (FPGA).  Analog sampling occurs continuously, while digitization and readout occur only when triggered.  The compact TARGET design is particularly well-suited to, but not limited to, the densely pixelated focal planes of IACTs using dual-mirror Schwarzchild-Couder~\cite{SchwarzschildCouder} optics. 

The ASIC design presented here is the natural evolution of a series of ASICs that have been developed for radio neutrino detection~\cite{STRAW,LAB3}, recording of photodetector output with precision timing~\cite{BLAB1,PrecTime}, and highly integrated photodetector readout~\cite{PROMPT}.  In order to meet the demands of next-generation IACT arrays for very-high-energy (VHE) gamma-ray astronomy, we draw upon this ASIC development experience to optimize the design of such highly integrated, cost-effective readout.  The TARGET ASIC can be compared with the ARS (Analogue Ring Sampler) ASIC~\cite{ARS2000,ARS2003} used for ANTARES and H.E.S.S.~I readout and the SAM (Swift Analogue Memory) ASIC~\cite{SAM} used for H.E.S.S.~II readout.  MAGIC uses DRS (Domino Ring Sampler) chips~\cite{Magic2Readout} and VERITAS uses flash ADCs~\cite{VeritasReadout}.

Considering both scientific drivers and budget constraints, the requirements for the camera electronics are as follows:

\begin{enumerate}
\renewcommand{\labelenumi}{\arabic{enumi}.}
\item Sampling rate per channel : $\geq1$~GSa/s 
\item Readout time per event: $< 20$~$\mu$s % Provided by Hiro 23 July 2010 
\item Trigger latency tolerance: $> 2$~$\mu$s 
\item Dynamic range: $> 8$~bits 
\item Cost (w/out photosensor): $<$ \$20/channel
\end{enumerate}

These design requirements are being met with the TARGET series of chips, as will be described below.  The first version of the chip (TARGET~1) meets most requirements but its dead time is somewhat too large.  This and other features will be improved with TARGET~2.  Section~\ref{sec:architecture} provides an overview of the TARGET~1 architecture including a description of the sampling buffer organization, self-triggering mechanism, and digitization methods.  Section~\ref{sec:test_system} describes the circuit board we used to evaluate the performance of the TARGET~1 chip.  Results of these tests are presented in Section~\ref{sec:tests}.  Section~\ref{sec:camera_module} describes a prototype of an IACT front-end readout camera module that we have produced.  Section~\ref{sec:TARGET2} presents the design of the next-generation chip in the series, TARGET~2.  We conclude with a summary in Section~\ref{sec:summary}.

Results from performance testing of the TARGET~1 chip, along with the expected performance of the TARGET~2 chip, are summarized in Table~\ref{specTable}.

%Future design goals and applications beyond TeV $\gamma$-ray astronomy are discussed in Section~\ref{sec:future}.  

%%%%%%%%%%%%%%%%%%%%%% Architectural %%%%%%%%%%%%%%%%%%%%%%%%%%%%%%%%%%%%

\section{Architecture}
\label{sec:architecture}

The TARGET ASIC has been designed to meet the requirements described in Section~\ref{sec:introduction}.  The key features of the TARGET design are its high-frequency (GSa/s) sampling, integrated trigger, high channel multiplicity, and deep buffer.  TARGET also features multi-hit buffering capability.  For multi-hit buffering, the buffer is divided in two and operated in ping-pong mode.  This decreases the dead time dramatically at the expense of halving the buffer depth.

% TARGET~2 parameters from Hiro's page
% https://confluence.slac.stanford.edu/display/AGIS/TARGET2+Specifications
% He says gain for bandwidth spec needs to be checked with UCSC
\begin{table*}[hbt]
\caption{\it Performance parameters of the first-generation chip (TARGET~1), as measured and reported here, as well as expected performance of the second-generation chip (TARGET~2).  In TARGET~1, both rising and falling edges of the Wilkinson ADC oscillator are used, in order to achieve a counter speed of 445~MHz with a clock speed of 222.5~MHz.  In TARGET~2, we use one edge per cycle with a faster clock, such that both the oscillator and counter speeds are 700~MHz.  The TARGET~2 analog bandwidth is specified with a gain of 60 and 50~$\Omega$ impedance.  In this table we define cross talk to be the ratio between digitized signals (i.e. after signal attenuation due to analog bandwidth and AC saturation).  In TARGET~2, cells must be digitized in blocks of 32 but a fractional number of blocks can be read out from the chip after digitization.  Dead time is expressed as digitization time + readout time.  Because the Wilkinson counters are implemented outside the ASIC (on an FPGA) in TARGET~1, the ASIC itself has zero readout time.  In TARGET~2, the counters are implemented inside the ASIC, so readout time is required to transfer to the FPGA.  Additional readout time may be incurred downstream of the FPGA with either chip but depends on the system design and can be small.  The readout time for TARGET~2 assumes we always transfer 12 bits per sample plus 3 bits of overhead for addressing, so 48 samples $\times$ 15 bits at 100 Mbps = 7.2~$\mu$s of readout time per event.  The number of bits given for each digitization time is the total dynamic range, without subtracting noise to give the effective dynamic range.}
\label{specTable}
 \begin{center}
    \begin{tabular}{|c|c|c|} \hline
      {\bf Parameter} & {\bf TARGET~1 } & {\bf TARGET~2} \\ \hline
     Channels & 16 & 16 \\ \hline
     Dynamic range (bits) & 9 or 10 & up to 12 \\ \hline
     Sampling frequency (GSa/s) & 0.7--2.3 & 0.2--1.8 \\ \hline
     3~dB analog bandwidth (MHz) & 150 & $>$~380 \\ \hline
     Cross talk at 3~dB frequency & $<$~4\% & 1\% \\ \hline    % this is measured wrt the digitized input signal, not the input signal, ie it is relative to the signal on the input channel after attenuation by the bandwidth function
     Buffer depth (cells per channel) & 4,096 & 16,384 \\ \hline
     Wilkinson ADC counter speed (MHz) & 445 & 700 \\ \hline
%     Simultaneous channels per digitization & 2 & 16 \\ \hline
     Samples per digitization (block size) & 16 & 32 \\ \hline
     Digitization time per block ($\mu$s) & 1 (9 bit) or 2 (10 bit) & 0.7 (9 bit) or 1.5 (10 bit) \\ \hline
     Number of Wilkinson ADCs & 32 & 512 \\ \hline
     Number of cells digitized simultaneously & 16 cells x 2 channels & 32 cells x 16 channels \\ \hline
     Clock speed for serial data transfer (Mbps) & - & 100 \\ \hline	% 100 Mbps = 100 MHz
     Channels for simultaneous data transfer & - & 16 \\ \hline
     Dead time for 48 samples $\times$ 16 ch ($\mu$s) & 24+0 (9 bit) or 48+0 (10 bit) & 1.5+7.2 (9 bit) or 2.9+7.2 (10 bit) \\ \hline
     % TARGET~2 was: 			                                                                                         & 1.5 + 8.64 (9 bit), 2.9 + 9.6 (10 bit) \\ \hline
     % Correction (Hiro email May 5 2011):  Addressing will have some overhead (3 cycles). So, 48*15/100=7.2

     Trigger outputs & 1 (OR of 16 channels) & 4 (each is analog sum of 4 channels) \\ \hline
     \end{tabular}
  \end{center}
\end{table*}   

\begin{figure}[htb]
\vspace*{0mm}
%\centerline{\psfig{file=input_coupling.eps,width=3.0in}}
\noindent\includegraphics[width=0.48\textwidth]{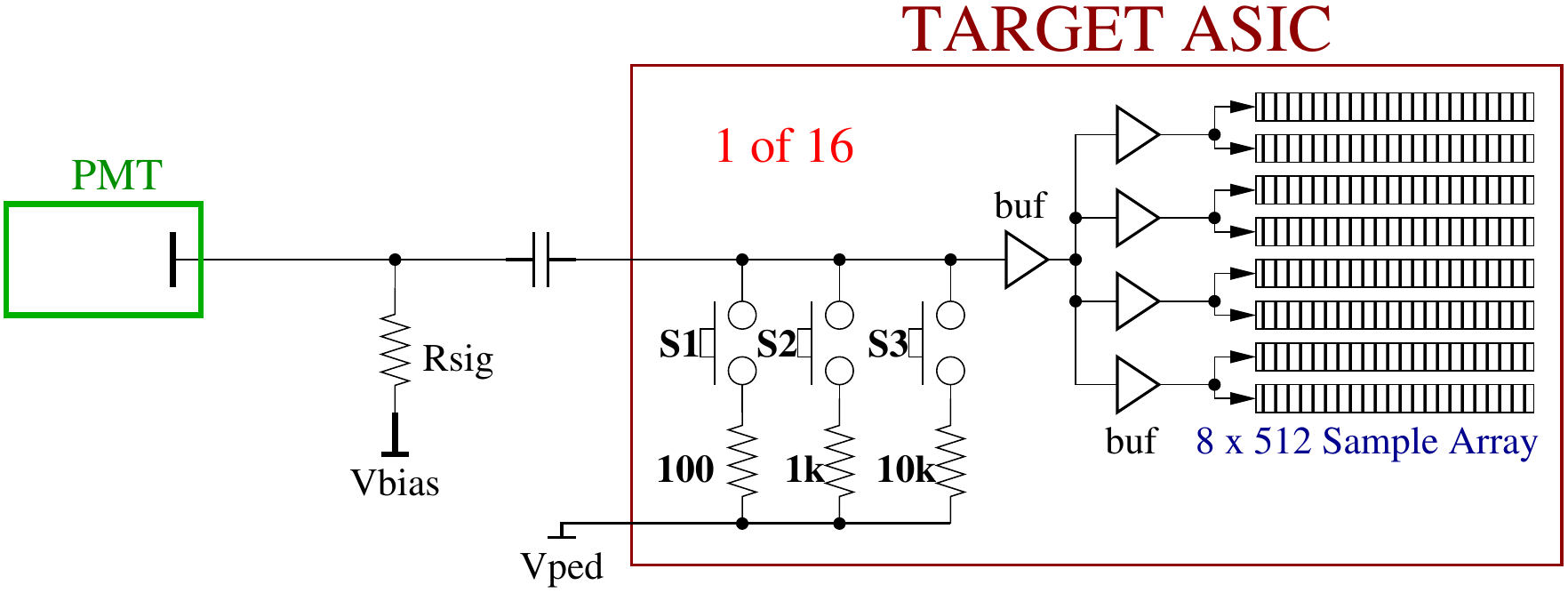}
\vspace*{0mm}
\caption{Block diagram of a single channel of the TARGET~1 ASIC, with the input coupling, termination, buffer amplifier tree, and switched capacitor storage array shown. Each TARGET chip consists of 16 such channels.}
\label{BLOCK}
\end{figure}

An overview of the waveform sampling structure for a single input channel is shown schematically in Figure~\ref{BLOCK}.  Each TARGET chip can record 16 photodetector channels.  Each channel has its own switched capacitor array consisting of 8 rows of 512 storage cells, for a total of 4096 storage samples per channel.  In total, TARGET~1 contains 65,536 analog storage cells.  A pedestal voltage ($V_{\rm{ped}}$) is used to provide a DC offset to the AC input signals, because the ADCs digitize positive voltages.  Centering $V_{\rm{ped}}$ in the middle of the ADC range allows the largest dynamic range for AC signals.

In typical usage, the TARGET chip is interfaced to an FPGA that provides all necessary configuration and control signals.  A single FPGA can control one or several TARGET chips depending on the channel multiplicity requirements of the application.  Firmware running on the FPGA in turn interfaces with data acquisition (DAQ) software on a computer.  This DAQ software configures the data taking parameters, starts and stops acquisition, and receives recorded events during acquisition.

%\begin{figure}[htb]

As shown in Figure~\ref{BLOCK}, each input channel has a variable input termination resistance.  This allows varying the voltage gain for a photodetector output that can be treated as a current source.  The switches shown in Figure~\ref{BLOCK} can be configured individually by firmware to set the input impedance.  Because directly driving the net capacitance of 4096 storage cells would significantly limit the analog input bandwidth, an analog buffer tree, consisting of unity gain buffers, is used as shown.

The capture window (number of samples digitized per waveform per channel) can be configured in firmware and must be a multiple of 16-sample blocks.  In testing we have operated with both three and four blocks per waveform (48 and 64 samples per waveform, respectively).

\begin{figure}[]	% htb
\vspace*{0mm}
%\centerline{\psfig{file=strobes.eps,width=3.0in}}
%\noindent\includegraphics[width=0.48\textwidth]{strobes.pdf}                             % old figure by Gary
\noindent\includegraphics[width=0.48\textwidth]{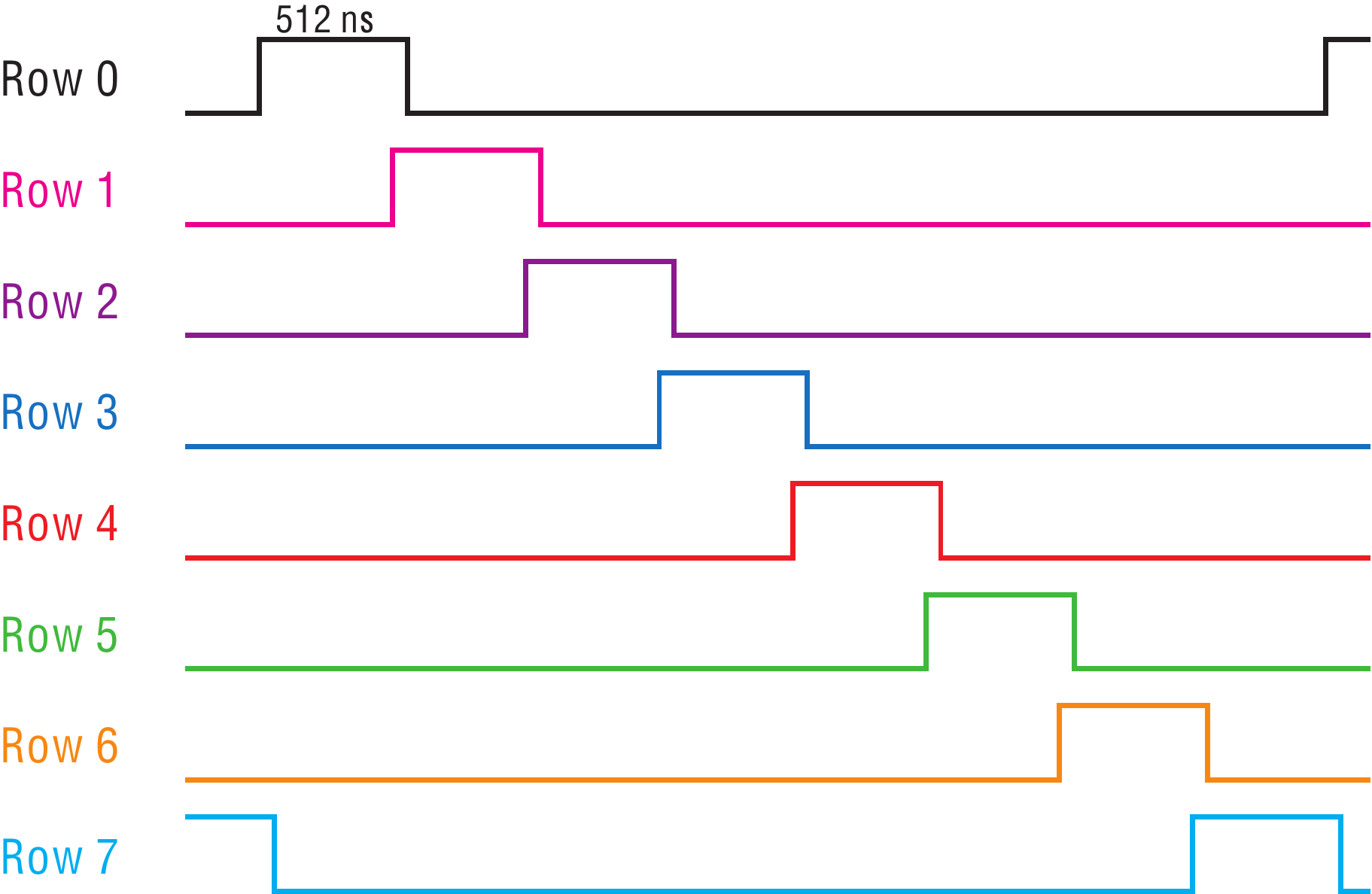} 		% new figure by Hiro
\vspace*{0mm}
\caption{A simplified diagram of the TARGET~1 sampling sequence.  Sampling proceeds through the 8 capacitor rows one by one.  The state machine that drives this sampling sequence is implemented in a companion FPGA.}
\label{strobes}
\end{figure}

Storing samples in each of the 8 rows of sample capacitors may be done quasi-independently.  Each row has a dedicated control line (``enable'' signal) to enable its sampling.  All even rows are connected to a common ``write strobe'' and all odd rows are connected to a separate write strobe.  Continuous sampling is achieved by alternating between driving the odd and even timing strobes and cycling through the row enable signals, as illustrated in Figure~\ref{strobes}.

\begin{figure*}[]		% htb
\vspace*{0mm}
\centering
%\centerline{\psfig{file=overblock.eps,width=3.0in}}
\noindent\includegraphics[width=0.95\textwidth]{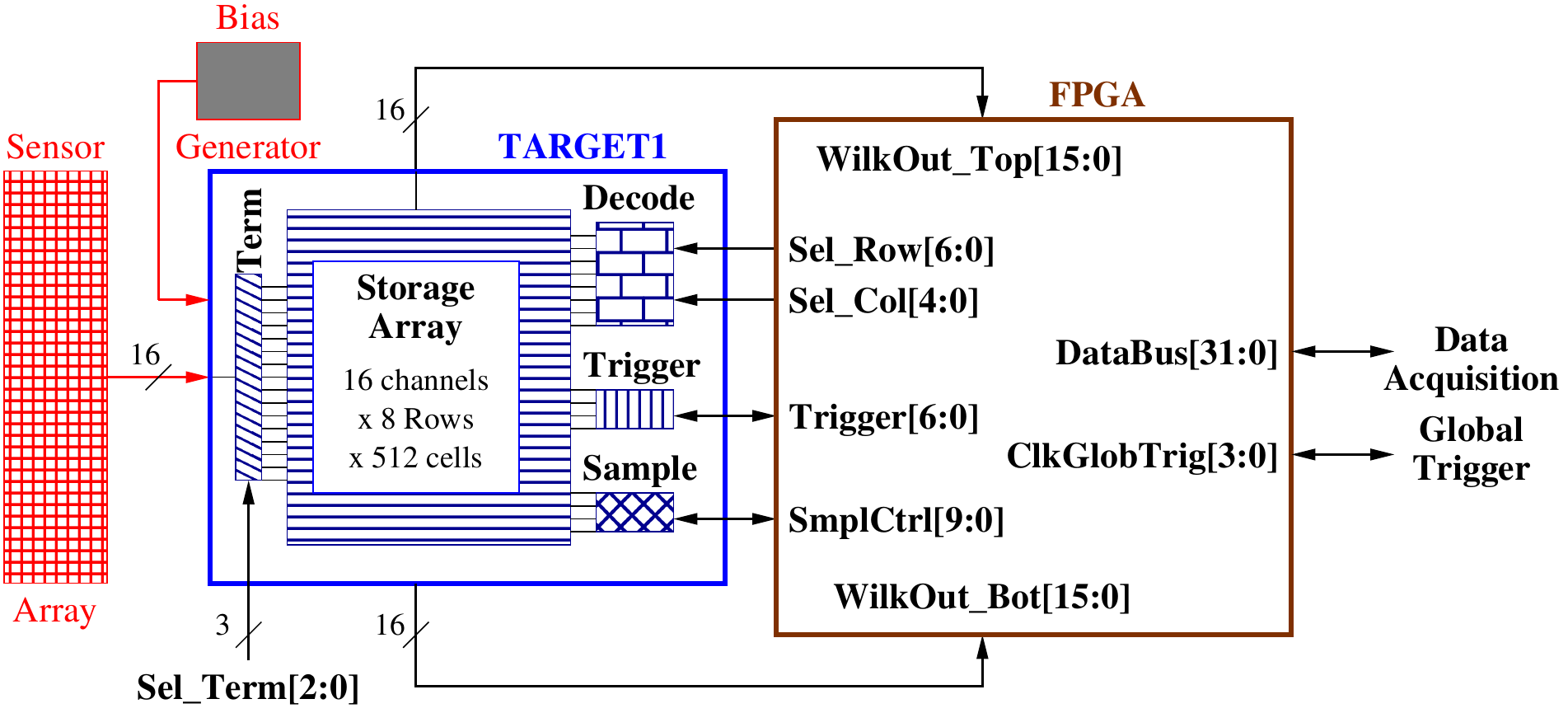}
\vspace*{0mm}
%\caption{test}
\caption{Functional TARGET~1 block diagram, with control signals and data paths indicated.  After generating an internal trigger (or receiving an external trigger), a group of 16 analog samples are selected from the first and second group of 8 channels, in parallel.  The ``enable'' and ``write strobe'' control lines are indicated as {\tt SmplCtrl} (8 enable signals and 2 write strobes).  The 8 enable signals correspond to the timing diagram in Figure~\ref{strobes}.  {\tt WilkOut\_Top} provides the comparator outputs from one bank of 16 Wilkinson ramps, and {\tt WilkOut\_Bot} provides the outputs from the other bank of 16.  The three control bits in {\tt Sel\_Term} control the three independent switches for selecting the input impedance.  The FPGA tells the TARGET~1 chip which blocks to digitize with {\tt Sel\_Row} (3 bits identify the row and 3 bits identify the pair of channels to read out simultaneously) and {\tt Sel\_Col} (5 bits identify the column).  The 7-bit {\tt Trigger} bus configures the trigger.}
\label{overblock}
\end{figure*}

Figure~\ref{overblock} shows a block diagram of TARGET~1 and its interfaces with a sensor array and FPGA in typical operation.  Each of the 16 input channels has a comparator circuit for self-triggering.  Each channel has a comparator and the OR (digital sum) of the individual comparator outputs provides a single global trigger.  An analog sum (multiplicity signal) is also available.  An external control voltage provides the analog comparator threshold for the trigger.  The same threshold is applied to all channels.  Future versions of the chip may allow some threshold variation from channel to channel, in order to compensate channel-to-channel gain variations in photo-sensors.  The polarity of the trigger output signal is configurable.  In addition to the internal trigger, waveform digitization and readout can also be initiated by an external trigger.  In an IACT camera, the external trigger will allow single-chip TARGET triggers to flow up to a high-level trigger decision followed by TARGET readout initiated by an external trigger flowing back to the TARGET chip.  The higher level trigger decision can be made either at the single-telescope or multiple-telescope level.  All channels are included in the trigger: there is currently no capability to mask out dead or noisy channels.  The width of the trigger signal is tunable.

%\begin{figure}[ht]
%\vspace*{0mm}
%\centerline{\psfig{file=TARGET_floorplan.eps,width=3.0in}}
%\vspace*{0mm}
%\caption{Floorplan diagram of the TARGET ASIC.  The die is 2.6 mm
%by x.yy mm and is fabricated in the TSMC 0.25$\mu $m process.}
%\label{floorplan}
%\end{figure}

%\clearpage

% These measurements and calculations are documented by Keith at
% https://confluence.slac.stanford.edu/display/AGIS/Digitization+and+Readout+Time

\begin{figure*}[]		% t
\centering
%\noindent\includegraphics[width=0.48\textwidth]{terminology}
\noindent\includegraphics[width=0.95\textwidth]{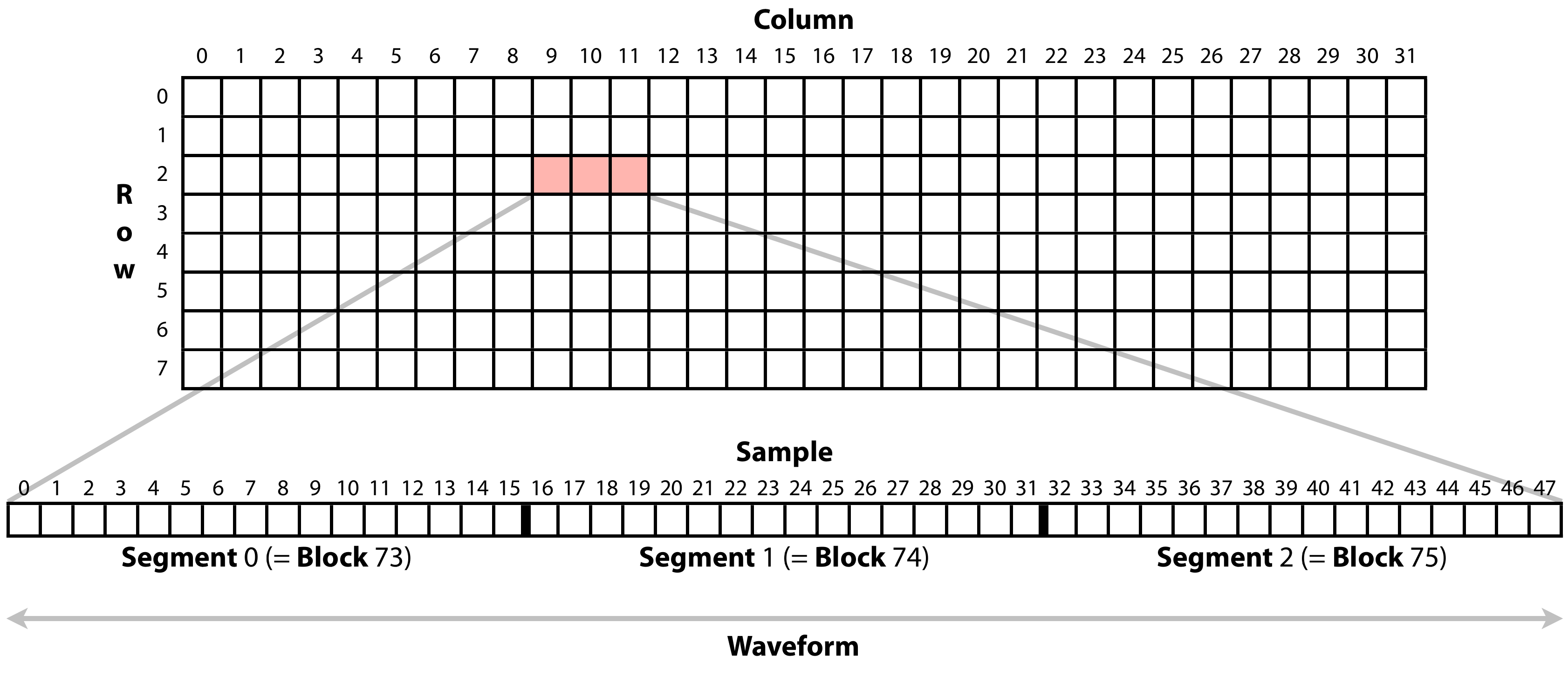}
\caption{Schematic diagram of the TARGET~1 switched capacitor array, including terminology used to describe TARGET sample acquisition.  The entire structure is repeated for each of 16 signal input channels per TARGET chip.  In this example, 3 blocks are read out to produce a waveform of 48 samples.}
\label{terminology}
\end{figure*}

The layout of the switched capacitor array and of the resulting waveform is shown in Figure~\ref{terminology}.  The TARGET~1 buffer consists of 4096 sample capacitors for each of the 16 signal input channels.  At 1~GSa/s, this provides a buffer depth of 4.096~$\mu$s for each channel.  The capacitors are arranged in 8 {\bf rows}, with 32 {\bf columns} per row.  Sample acquisition proceeds from one column to the next within each row, and from one row to the next when each row is filled.  Each row-column combination specifies a {\bf block} of 16 capacitors.  Within each block, sample acquisition proceeds from one capacitor to the next.  When a trigger occurs, several consecutive blocks are digitized and read out to produce a waveform.

The number of blocks read out per event can be specified in firmware.  Within a waveform, the blocks of samples are referred to as {\bf segments}.  While the term ``block'' refers to a physical set of capacitors in the storage array, the term ``segment'' refers to a set of 16 consecutive samples within a waveform, without regard to which physical set of capacitors acquired them.  Some chip behaviors vary according to block ID and some vary according to segment ID.  All quantities are indexed from zero.  For most of the tests reported here, 4 blocks (64 samples) were read out per waveform.

Waveform digitization begins with the arrival of a trigger signal.  Digitization is performed with Wilkinson analog-to-digital converters (ADCs).  The Wilkinson ramp works as follows: a ramp voltage is increased linearly until it equals the sample capacitor voltage.  A 12-bit digital gray-code counter is started when the ramp starts.  When the ramp voltage matches the capacitor voltage, a comparator stops the digital counter.  The value of the digital counter then provides the ADC code corresponding to the input voltage.  In the TARGET~1 design, the Wilkinson ramps and comparators are implemented in the TARGET ASIC and the counters are implemented in an FPGA.  The comparator outputs are routed from the TARGET~1 ASIC to the FPGA to stop the counters.

Each TARGET~1 ASIC has two banks of 16 Wilkinson ramps (one for channels 0--7 and one for channels 8--15), so that 32 waveform samples are digitized simultaneously.  Accordingly, blocks of 16 consecutive cells in the storage buffer form the basic unit of the digitization window selection.  Individual blocks may be randomly accessed.  Blocks from two distinct channels are digitized in parallel (channels 0 and 8, channels 1 and 9, etc.) by the two sets of Wilkinson ramps.

In the evaluation board used for the tests reported here, Wilkinson counting is accomplished using a clock signal that is 222.5~MHz and counting on both rising and falling edges of the clock signal in order to achieve a 445~MHz counter.  While the counter speed is fixed, the Wilkinson ramp speed can be adjusted to achieve the desired configuration, with a tradeoff between dynamic range and digitization time.  In the configuration used for the tests described below, the time to complete the full 12-bit (4096-step) range of each Wilkinson ramp is 9.2 $\mu$s.  The digitization time (ramp time) can be divided in half by digitizing 11 bits instead of 12 bits, in four by digitizing 10 bits, and so on.

A trigger results in digitization and readout of all 16 channels.  A typical selection window to capture Cherenkov pulses from IACTs at 1 GSa/s is 64 samples, corresponding to 4 digitization blocks from each input channel.  In the configuration used for the tests below, the total time to digitize 4 blocks in each of 8 channel pairs is $4 \times 8 \times 9.2 = 294$~$\mu$s.  Faster readout time can be achieved by decreasing the dynamic range from 12 to 10 or 9 bits.

In addition to the standard mode described here, TARGET~1 could be operated in a  multi-hit mode.  The 4096-sample buffer would be operated as two 2048-sample buffers.  While digitization and readout of one buffer are occurring, the other buffer would continue sampling and ping-pong operation would proceed back and forth between the two.  This mode would be valuable for applications that require higher readout rates and/or lower dead time and can accept a smaller buffer depth.

%%%%%%%%%%%%%%%%%%%%%% Readout Test System %%%%%%%%%%%%%%%%%%%%%%%%%%%%%%%%%%%%

\section{Evaluation board}
\label{sec:test_system}

\begin{figure}[]	% ht
\vspace*{0mm}
%\noindent\includegraphics[width=0.48\textwidth]{TARGET_evall_pcb}
%\centerline{\psfig{file=TARGET_evall_pcb.eps,width=3.2in}}
%\noindent\includegraphics[width=0.37\textwidth, angle=-90]{EvalBoard.pdf}
\noindent\includegraphics[width=0.48\textwidth]{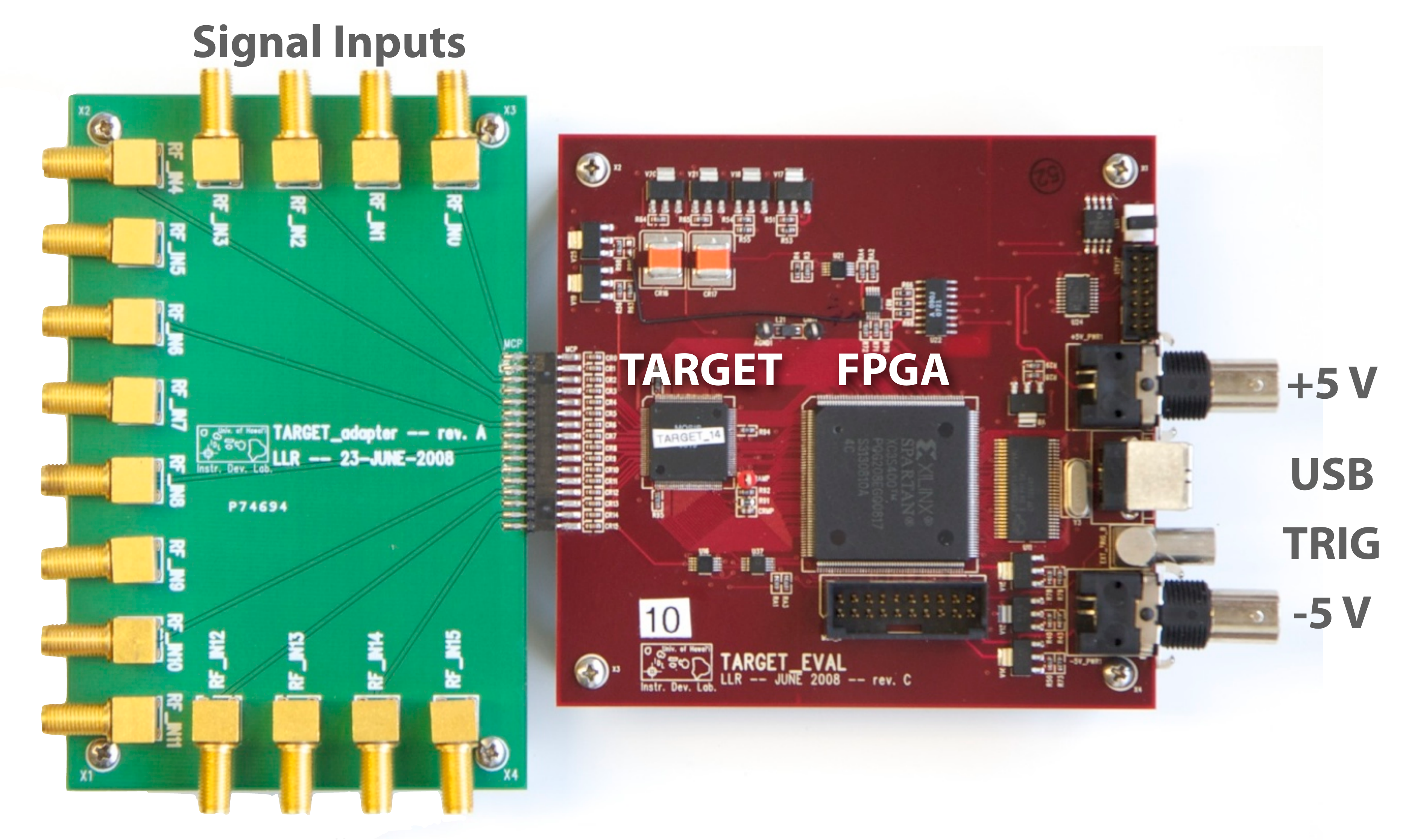}
\vspace*{0mm}
\caption{Photograph of the TARGET~1 evaluation board (right) and RF input board (left).  The 2-mm-pitch, dual-row connector is designed for direct interconnect to a 64-channel multi-anode photomultiplier tube (MAPMT) and also matches the RF input board.  The evaluation board includes one TARGET~1 chip, one FPGA, a USB interface, an external trigger link, and BNC connectors for +5~V and -5~V power.}
\label{TARGET_eval_PCBs}
\end{figure}

Evaluation of the TARGET~1 ASIC has been performed using the circuit boards shown in Figure~\ref{TARGET_eval_PCBs}.  On the left is a board that permits the insertion of high-frequency signals via SMA input connectors.  This board connects to a board (hereafter ``evaluation board'') featuring one TARGET~1 chip and all other components necessary to support it.

The three main components on the evaluation board are a single TARGET~1 ASIC (in a 120-pin thin quad flat pack chip), an FPGA, and a USB interface.  The external communication is via USB 2.0, achieved with a Cypress CY7C68013-56PVC controller.  The FPGA (Xilinx Spartan XC3S400) controls the digital logic and timing for TARGET~1 readout.  Internal FPGA RAM buffers the data before transfer to the computer via USB.  DAQ software running on a computer configures the FPGA through its firmware and receives the TARGET data.  Both occur via the USB 2.0 interface.  The evaluation board interfaces available are: 16 signal pins, +5~V and -5V power, external trigger (a board input used to trigger TARGET by an external source), USB, JTAG for installing new firmware, and a block of pins to monitor digital lines.

%%%%%%%%%%%%%%%%%%%%%% Readout Test System %%%%%%%%%%%%%%%%%%%%%%%%%%%%%%%%%%%%

\section{Test results}
\label{sec:tests}

We used the evaluation board to determine the basic performance parameters of the TARGET~1 chip.  Results are reported below.

% https://confluence.slac.stanford.edu/display/AGIS/February+2010
\subsection{Power consumption}

We measured the power consumption of the TARGET~1 chip alone to be 113~mW (7~mW per channel) at room temperature and a sampling frequency of 1~GHz.  Most (100~mW) of the power consumed by the chip is due to front-end buffer amplifiers, whose consumption could be reduced if necessary.  Increasing the event readout rate from 0~Hz to 70~Hz increases the chip power consumption by 20~mW.  The power consumption of the entire evaluation board (including the FPGA that reads out the chip) is 1.7~W at 0~Hz and increases to 1.8~W at 70~Hz.  The camera module prototype (which includes 4 TARGET~1 chips and 1 FPGA and is described in Section~\ref{sec:camera_module}) consumes 4.29~W at 100~Hz trigger rate, increasing linearly with trigger rate to 4.35~W at 3.3~kHz (the maximum rate currently supported with the fiber optic interface).
%No significant variation in the TARGET chip power consumption was found across a range of operating temperatures and sampling frequencies. {\bf Check these statements by direct measurement.}

%{\bf TODO: }
%\vspace{0.05in}
%\begin{itemize}
%\item Temperature - probably no variation  
%\item Sampling frequency - probably weak dependence
%\item Readout rate - we expect power to go linearly
%\item Contribution of buffer amps to quiescent current?
%\end{itemize}

\subsection{Sampling frequency}

\begin{figure}[]
\vspace*{0mm}
\noindent\includegraphics[width=0.48\textwidth]{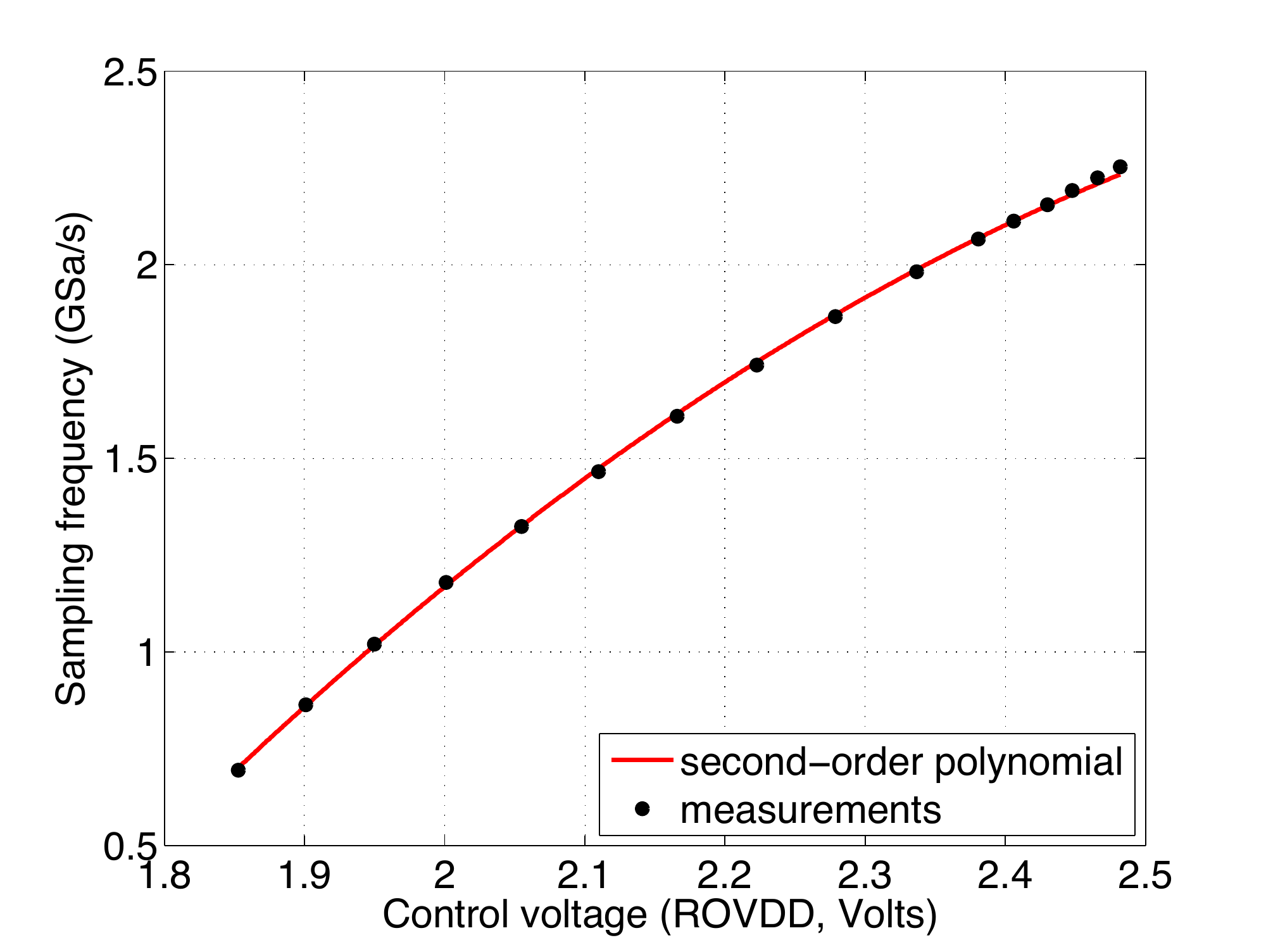}
\vspace*{0mm}
\caption{Measured sampling frequency as a function of control voltage (ROVDD).  This measurement was made with row 4 and columns 16 through 19, at room temperature.  The normal operating range is 0.7~GSa/s to 2.3~GSa/s.}
% rows and columns are indexed from 0
\label{rovdd}
\end{figure}

The sampling frequency of the chip is determined by two external voltages (ROVDD and ROGND) that together steer a voltage-controlled delay line.  The sampling frequency of the chip was measured at room temperature by recording a 60~MHz, 500~mV peak-to-peak sine wave.  A sinusoid was then fit to each recorded waveform to determine the sampling frequency as a function of ROVDD, as shown in Figure~\ref{rovdd}.  Sampling frequencies between 0.7~GSa/s and 2.3~GSa/s are possible.  

%\begin{figure}[ht]

Without temperature compensation, the sampling frequency will vary for a fixed ROVDD voltage.  In practice, the sampling frequency is held steady through the use of a feedback loop applied by the companion FPGA.  The TARGET~1 chip produces a ripple oscillator output, proportional to sampling speed, whose frequency can be compared to that of an external clock.  ROVDD is adjusted by the FPGA via an external digital-to-analog converter (DAC) to lock the desired sampling frequency.

\begin{figure}[]
\centering
%\noindent\includegraphics[width=0.48\textwidth]{rate_vs_temp_stat}
\noindent\includegraphics[width=0.48\textwidth]{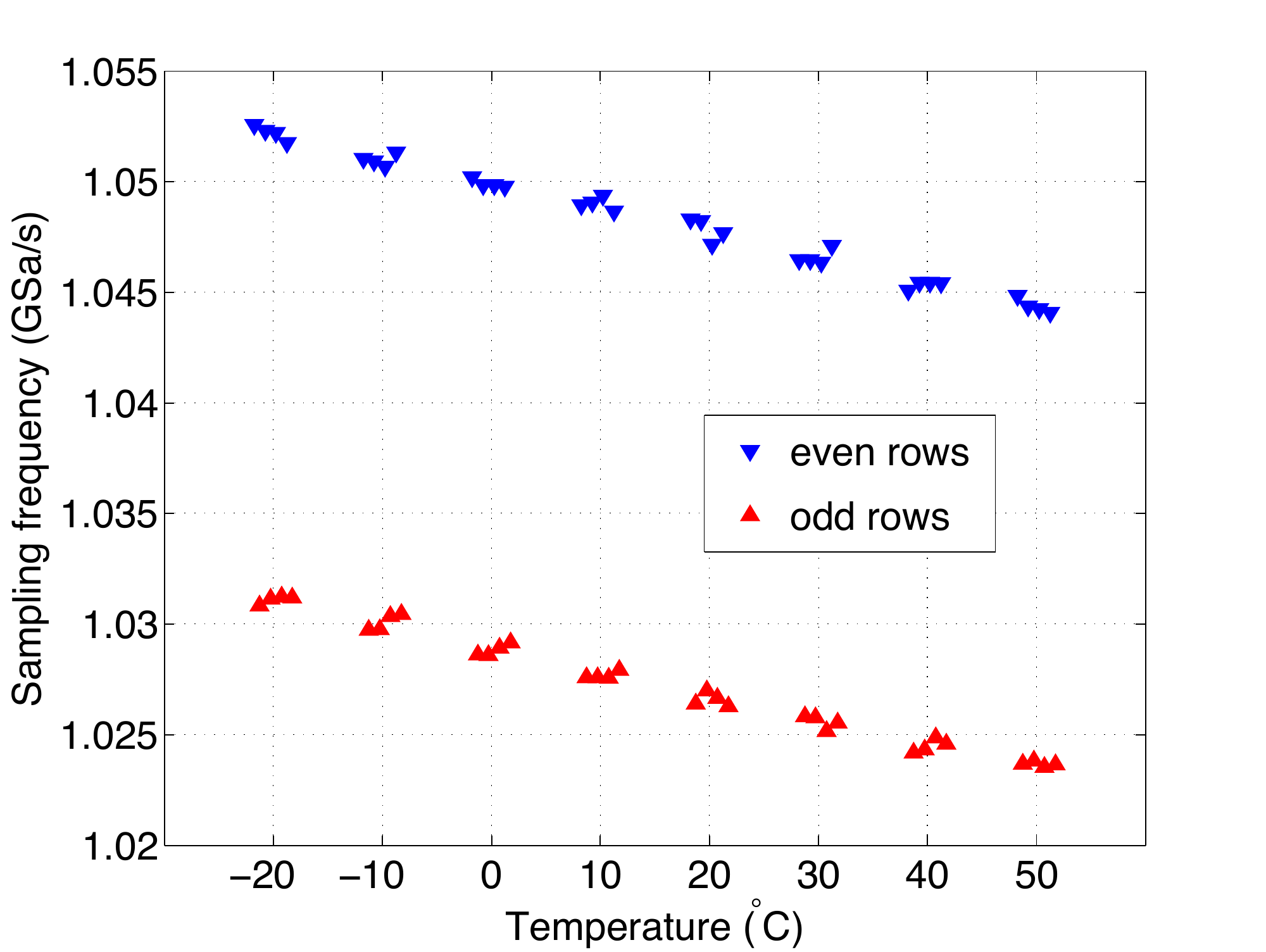}
\caption{Sampling frequency vs. temperature, using the control loop described in the text.  Measurements were made every 10.0~$^\circ$C but a small row-dependent horizontal offset has been applied in the plot for clarity.  The nominal sampling frequency for this test was 1~GSa/s, selected with an ROVDD value of 1.955~V and a ROGND value of 0.595~V (these were the default values used for most of our tests).  The sampling frequency is weakly dependent on temperature, decreasing by 0.7\% from -20~$^{\circ}$C to +50~$^{\circ}$C.  There is larger intrinsic difference between sampling rows: the four even rows sample 2.0\% faster than the four odd rows.}
\label{rate_vs_temp}
\end{figure}

To determine the performance of this feedback loop, the evaluation board was tested in a thermal chamber (TestEquity model 1007C).  The temperature was varied between -20 and +50~$^{\circ}$C while recording with a nominal sampling frequency of 1~GSa/s.  A cable feed-through into the thermal chamber allowed the evaluation board to be in the thermal chamber while all other electronics including power supplies and signal generator were outside of the thermal chamber, in order to isolate the temperature dependence of the TARGET~1 system itself.  The results of this measurement are shown in Figure~\ref{rate_vs_temp}.  The sampling frequency was found to vary weakly with temperature at a level which is consistent with the thermal stability of the clock on the evaluation board. Further precision in sampling frequency control could be achieved using an external reference clock.  

In addition to the small temperature dependence of the sampling frequency, there is small row-to-row variation.  The four even rows sample 2.0\% faster than the four odd rows.  This is due to necessary differences in the physical layout of the pair of sample timing generator circuits.  Within each row, however, the sampling frequency is stable.  The one-sigma variation of sampling frequency among all blocks in even rows is 0.23\%.  The variation among all blocks in odd rows is 0.13\%.  These sub-percent variations are negligible.  The 2.0\% variation between even and odd rows is also very small and should introduce no issues in photo-sensor readout.  Moreover it is deterministic and could be calibrated out by using a row-dependent sampling frequency to determine the absolute time of each sample.

%\begin{figure}[t]

% Keith's eps plot taken from https://confluence.slac.stanford.edu/display/AGIS/May+2010
% Justin's estimate of slope, from bottom of two curves:
% 50 C data point is 1.024 GHz
% -20 C data point is 1.031 GHz
% delta frequency is 1.031-1.024 = 0.007 GHz
% mean frequency is 1.0275 GHz
% so full range is 0.7%
% Variation from even to odd rows: example at -20 C:
% 1.031 GHz or 1.052 GHz:
% 2.0%
% TODO double check that frequency was calculated correctly from fitting to sinusoids (the logic of converting fitted to actual - Keith and I checked this once but should double check).

%{\bf This is the result for TARGET??}
%Determination of the sampling speed is made by measuring the time
%interval between insertion of the timing strobe and appearance of the
%output pulse from the last cell of the row, minus pad buffer
%delays. The sampling speed is calculated by taking the number of cells
%in a row and dividing it by the propagation time for a given control
%voltage setting. A plot of the sampling speed versus control voltage
%(ROVDD) is shown in Figure~\ref{rovdd}, where it is seen that sampling
%rates from below 0.3 GSa/s to above 3.5 GSa/s are possible.

% figure was here

\subsection{Trigger performance}

%\begin{figure}[t]
\begin{figure}[]
\centering
%\noindent\includegraphics[width=0.48\textwidth]{trigger_width_vs_temp}
\noindent\includegraphics[width=0.48\textwidth]{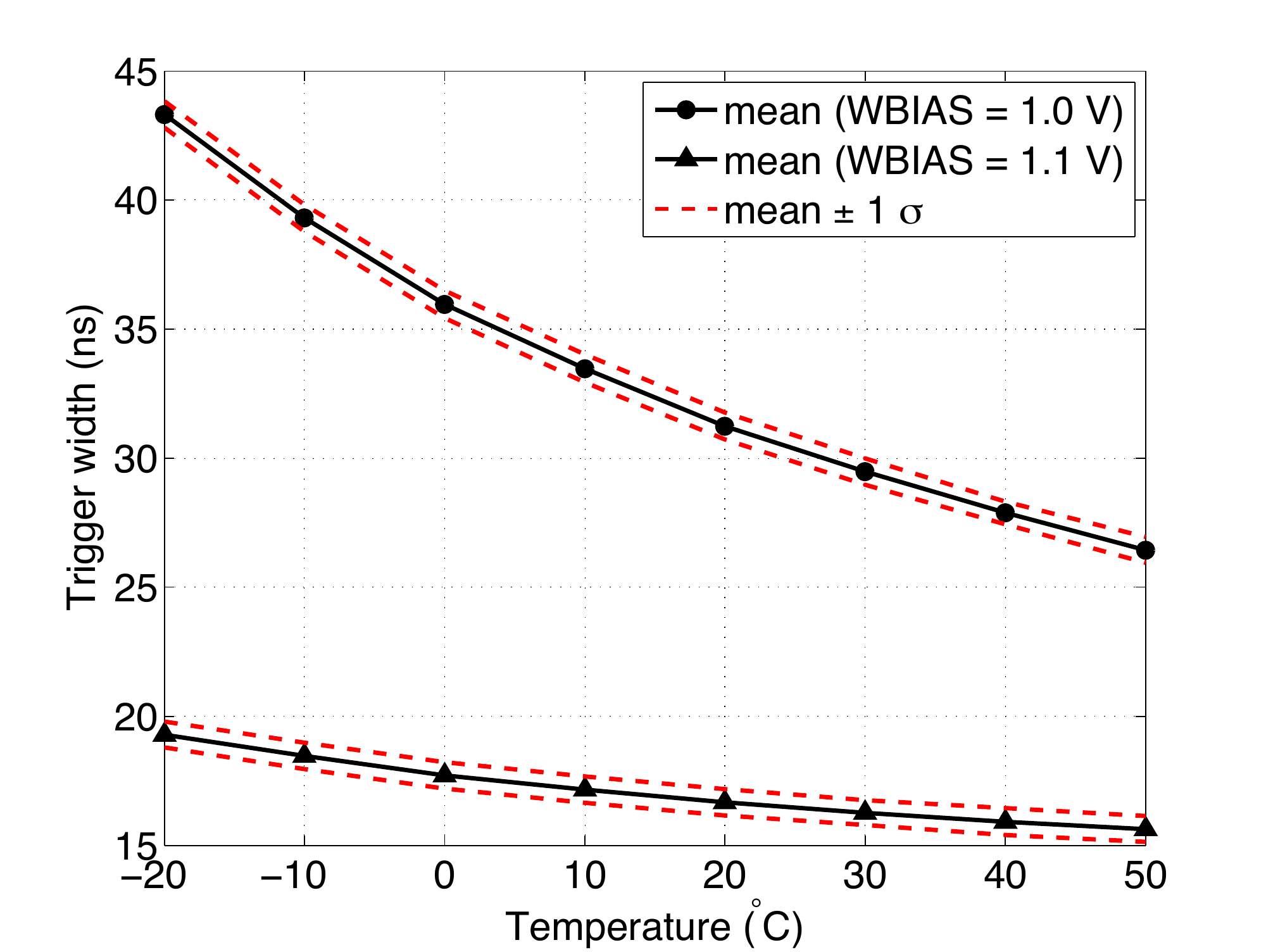}
\caption{Trigger signal width as a function of temperature.  The width can be tuned with a control voltage (WBIAS), and we measured the width vs. temperature for WBIAS values of 1.0 and 1.1~V.  The pulse-to-pulse variation was determined by measuring the standard deviation of 20,000 pulses at each temperature.}
\label{trigger_width_vs_temp}
\end{figure}

Because telescope ambient temperatures could vary significantly and the trigger circuitry is known to be temperature dependent, a test of the trigger performance vs. temperature was performed.  The evaluation board was operated in the thermal chamber between -20~$^\circ$C and +50~$^\circ$C.  After each temperature change, the board was allowed to equilibrate for five minutes before taking data.  A function generator was used to provide square signal pulses (70~Hz repetition rate, +200 mV pulse amplitude, 5~ns rise time, 8~ns high time, 5~ns fall time) and a digital oscilloscope was used to measure the width of the trigger (``DHIT'') signal for 20,000 events at each temperature.  From these events the mean and standard deviation of the trigger width were determined at each temperature.  The results of this measurement are shown in Figure~\ref{trigger_width_vs_temp}, for two values of the control voltage (WBIAS) that sets the trigger width.

These results indicate first that the trigger performs normally over the full temperature range and second that the trigger width is somewhat temperature dependent.  Other circuitry (such as higher-level single-telescope and multi-telescope triggers) that take the trigger signal as input may depend on the trigger width being within a certain range.  The measured width variation is small enough that such circuitry should be able to accommodate it.  If necessary, however, a control loop could be implemented in firmware to stabilize the trigger width.

% A feedback loop can be used to maintain a fixed trigger width.
% Commented out because actually this seems difficult to do - trigger width is very small compared to the ROVDD control loop where we count many cycles.

% plotTriggerEfficiency.m
%\begin{figure}[t]
\begin{figure}[]
\centering
\noindent\includegraphics[width=0.48\textwidth]{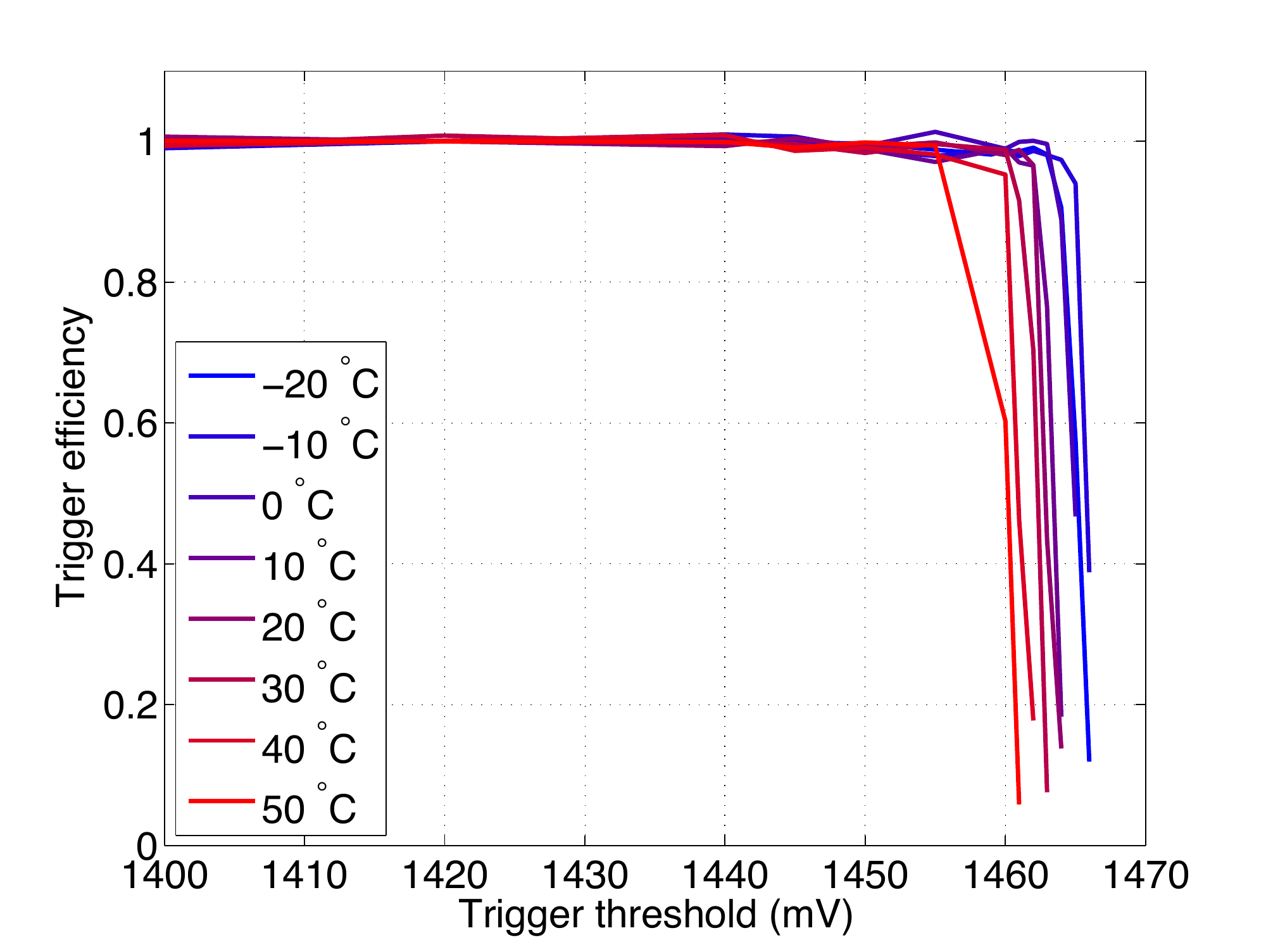}
\caption{Trigger efficiency vs. trigger threshold, for temperatures between -20~$^\circ$C and +50~$^\circ$C.  The efficiency curve varies little with temperature.}
\label{trigger_efficiency}
\end{figure}

Figure~\ref{trigger_efficiency} shows the TARGET~1 trigger efficiency vs. trigger threshold.  The chip triggers when the measured amplitude (after pedestal subtraction) exceeds the threshold.  It can be configured to trigger on rising or falling edges.  For this test, the pedestal voltage was $\sim$1400~mV and the signal was a square pulse with $\sim$100~mV amplitude.  The trigger efficiency vs. threshold was measured at temperatures between -20~$^\circ$C and +50~$^\circ$C.  The 50\% point (i.e., the threshold at which the trigger efficiency crosses 50\%) varied negligibly with temperature: it was between 1460 and 1467~mV for all eight temperatures tested.

\begin{figure*}
\begin{center}
\subfigure[]{
\label{mean_transfer_function}
\noindent\includegraphics[width=0.48\textwidth]{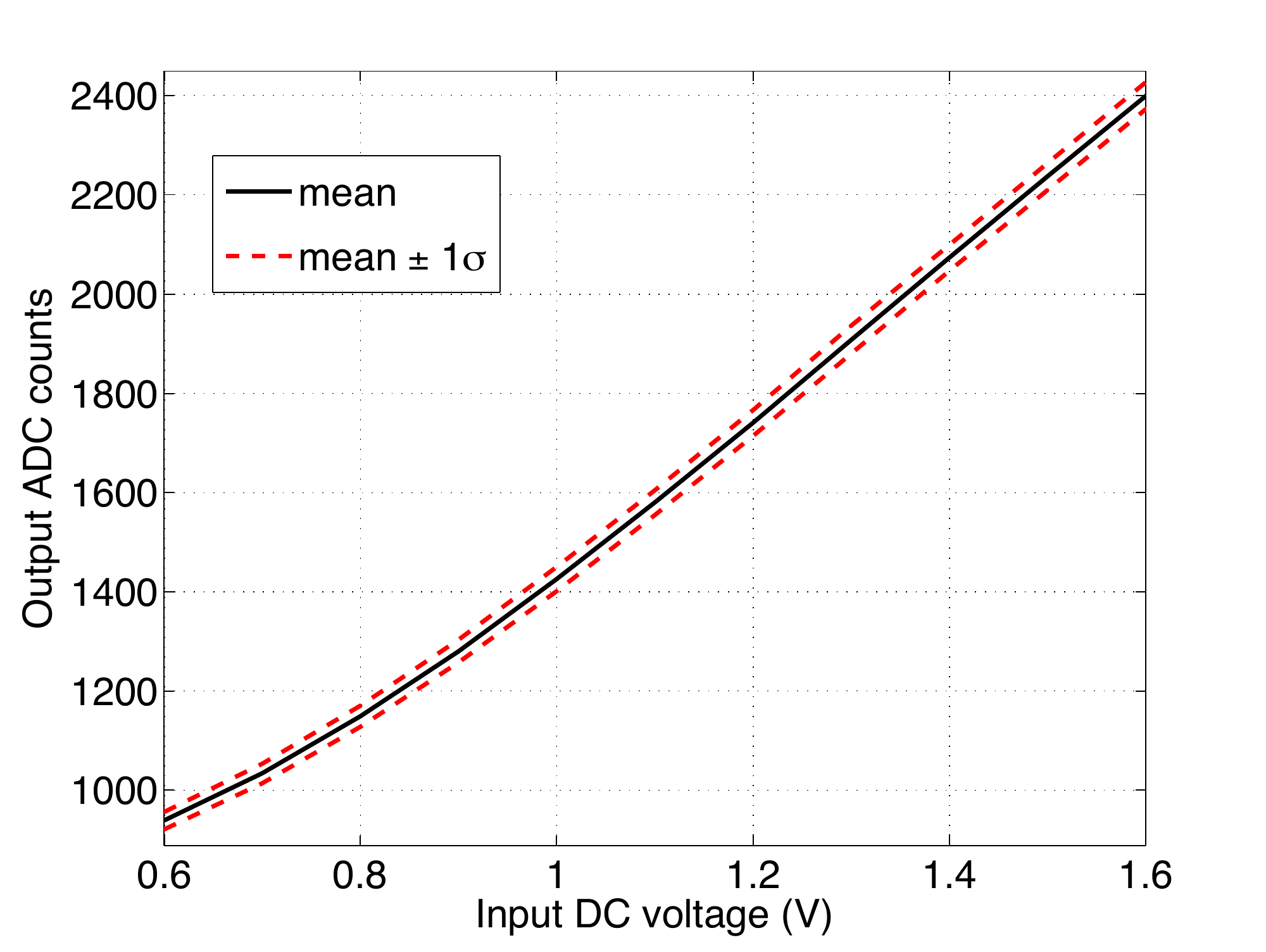}
}
\subfigure[]{
\label{function}
%\noindent\includegraphics[width=0.48\textwidth]{fitEachCapacitor_cap55.eps}
\noindent\includegraphics[width=0.48\textwidth]{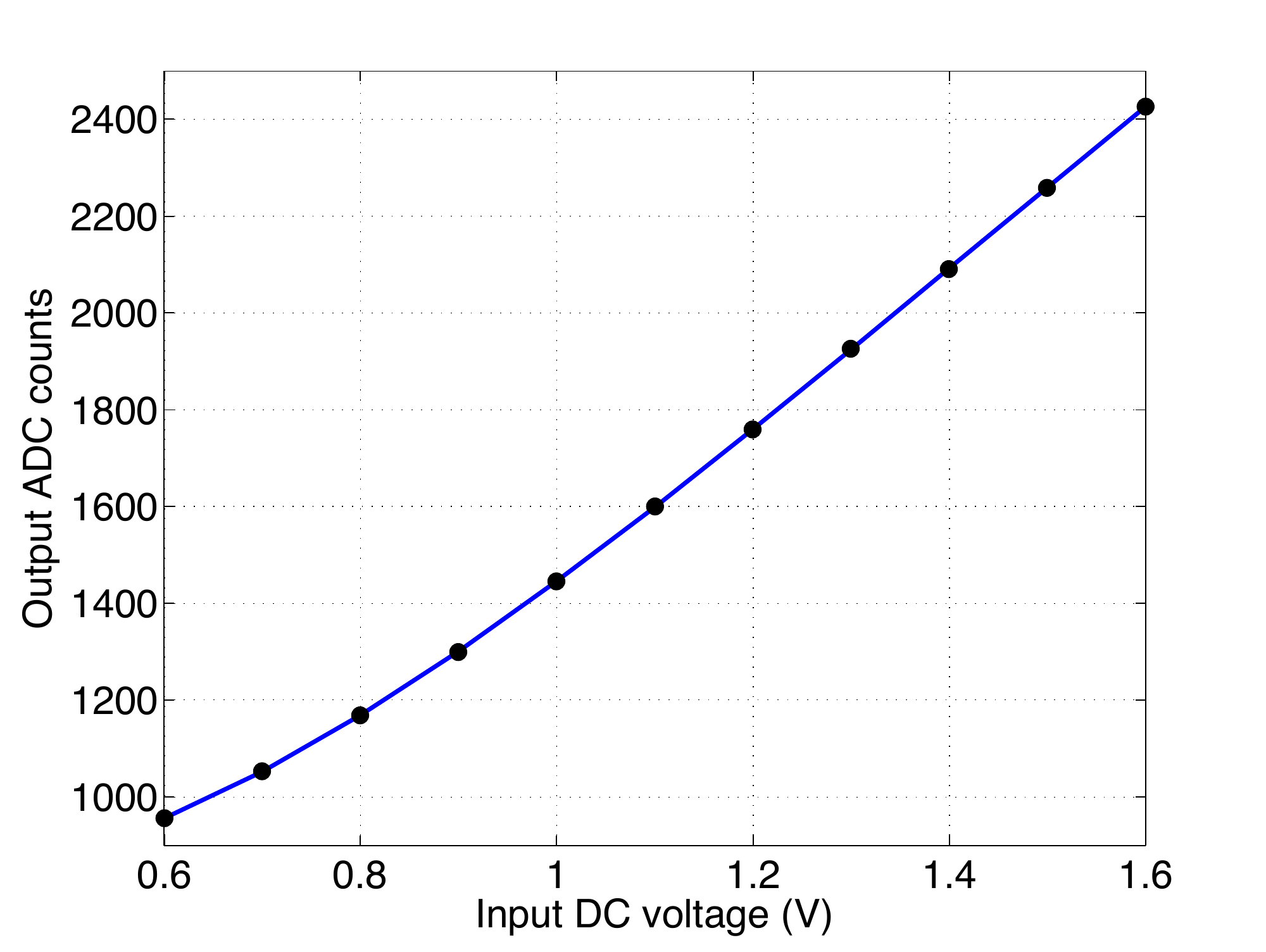}
}
\subfigure[]{
\label{residuals}
%\noindent\includegraphics[width=0.48\textwidth]{fitEachCapacitor_cap55_residuals}
\noindent\includegraphics[width=0.48\textwidth]{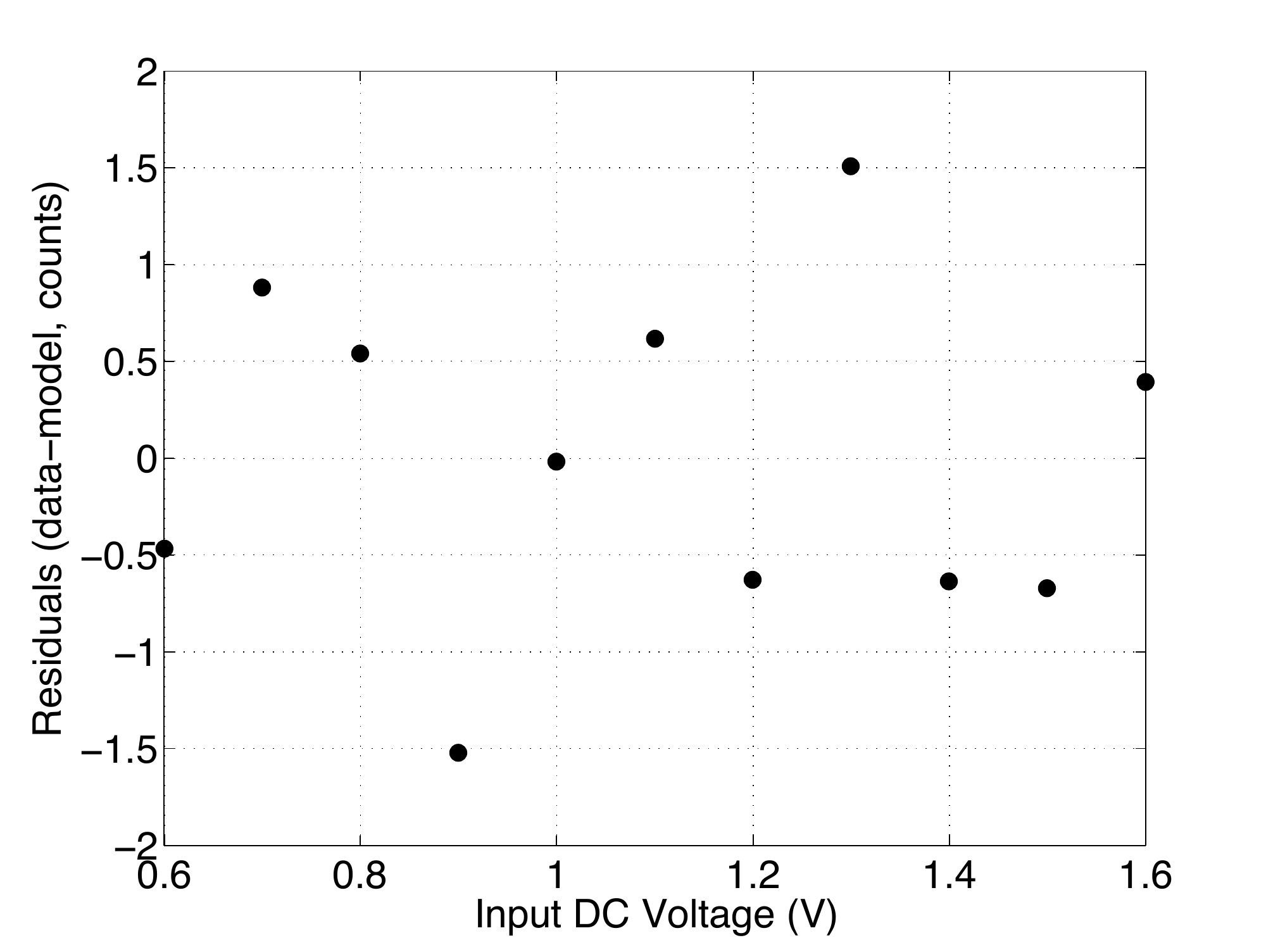}
}
\label{single_capacitor_transfer_function}
\caption{\subref{mean_transfer_function} Mean and variation of the transfer function for all 4096 capacitors composing the sampling array of a single TARGET~1 ASIC channel.  The transfer function was measured for each capacitor and the mean and standard deviation were determined from the ensemble of transfer functions.  The nominal operating range is 0.6 to 1.6~V.  \subref{function} Measured transfer function (points), along with fourth-order polynomial parameterization (curve), for a typical individual capacitor (Capacitor ID 55).  \subref{residuals}  Residuals of the parameterization shown in \subref{function}, for the same example capacitor.  The standard deviation of the residuals is 0.9 ADC counts for this capacitor.  This analysis includes all samples for each capacitor, without regard to segment position.  More precise calibration can be achieved by treating each segment position independently.}
\end{center}
\end{figure*}

\subsection{DC transfer function}
\label{transferFunction}

For a given direct-current (DC) input voltage, the chip outputs a particular 12-bit ADC code specifying the number of ADC counts recorded by the Wilkinson counter.  The transfer function mapping input DC voltage to output ADC counts is smooth and monotonic in the 0.6 to 1.6~V input range.  Although each capacitor in the storage buffer has a distinct transfer function, the variation among capacitors is small, as indicated in Figure~\ref{single_capacitor_transfer_function}.

% Gain value calculated from plotOurTransferFunction.m, agis_offline
Because the transfer function is not linear, the gain is a function of amplitude.  To determine an overall estimate of the gain, we fit the mean transfer function (combining all samples without regard to capacitor ID or segment ID) with a line in the 0.6 to 1.6~V range.  The slope of this best-fit line is the nominal ADC gain: 0.681~mV per ADC count.
%This is for a configuration with a resistor value of XXX, a ramp speed parameter of XXX.  This configuration has XXX counts (XXX bits) of dynamic range and a digitization time of XXX us per ramp.
% The newest code gives slightly different value, not 0.681, why?
% I believe this is because old code combines all segment and new code includes first segment only.
% I believe the results presented here are combining data from all segments.
% These results are for Board 03 Runs 127-147 recorded in Aug 2010, I think.

While the variation among capacitors is small, for precise waveform measurement it is worthwhile to calibrate each capacitor individually and also to calibrate each capacitor separately for each segment position in which it is read out.

We fit a transfer function to each of the 4096 capacitors of a single channel and analyzed the fit residuals.  Using a linear fit, the absolute value of the residuals is up to 122 ADC counts for some capacitors and some input voltages.  The mean standard deviation of the residuals (averaged over all capacitors) is 37.9 ADC counts using the linear parameterization for each capacitor.  The conclusion is that the most precise calibration can be achieved by using capacitor-dependent and non-linear (but smooth and monotonic) transfer functions.

A good fit to each capacitor's transfer function is achieved with a fourth-order polynomial parameterization.  With this parameterization, the absolute value of the residuals is less than 8.7 ADC counts for all capacitors and for all input voltages between 0.6 and 1.6~V (with one measurement every 0.1~V).  The mean standard deviation of the residuals (averaged over all capacitors of one channel) is 1.1 ADC count.  The best-fit transfer function and residuals for a single example capacitor are shown in Figure~\ref{single_capacitor_transfer_function}.  For a large-scale IACT array, a calibrated voltage source (digital-to-analog converter) can be built into each camera module and used for fast, automated calibration of transfer functions.

%\begin{figure}[t]
\begin{figure}[]
\centering
\noindent\includegraphics[width=0.48\textwidth]{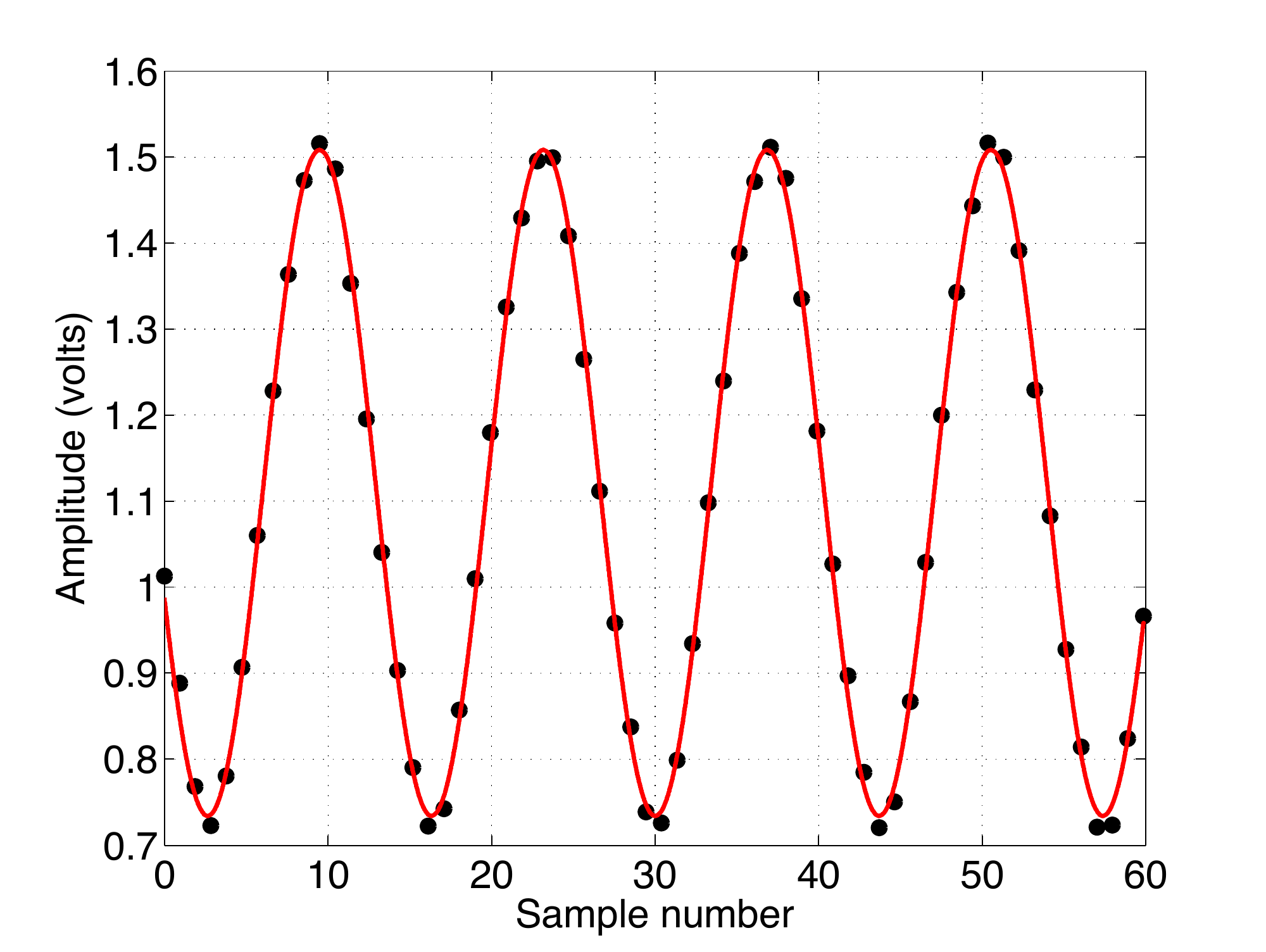}
% Figure generated with
% runFitSinusoidsOneRun(2199,1e3,990,73.1e6,1,8)
\caption{Calibrated sinusoidal waveform recorded by a TARGET~1 chip.  Recorded samples are plotted as dots and a sinusoidal fit is plotted as a curve.  The signal frequency was 73.1 MHz and the sampling frequency was 1.05 GSa/s.}
\label{waveform}
\end{figure}

The analysis described above includes all sample recordings from each capacitor, without regard to segment position.  More precise calibration can be achieved by treating each segment position separately.  Figure~\ref{waveform} shows an example calibrated waveform recorded with TARGET~1.  Capacitor- and segment- dependent transfer functions have been used to convert raw counts to voltages.

These transfer function results are for a particular Wilkinson ramp speed and corresponding results have been achieved for other ramp speeds.  The ramp speed can be tuned to achieve a larger dynamic range or smaller dead time.

%\begin{figure}[ht]
%\vspace*{0mm}
%\centerline{\psfig{file=TARGET_DNL_raw.eps,width=3.0in}}
%\vspace*{0mm}
%\caption{Determined output code as a function of DC input voltage.}
%\label{TARGET_DNL_raw}
%\end{figure}

\begin{figure*}
\begin{center}
\subfigure[]{
\label{overlaySaturation}
\centering
\noindent\includegraphics[width=0.48\textwidth]{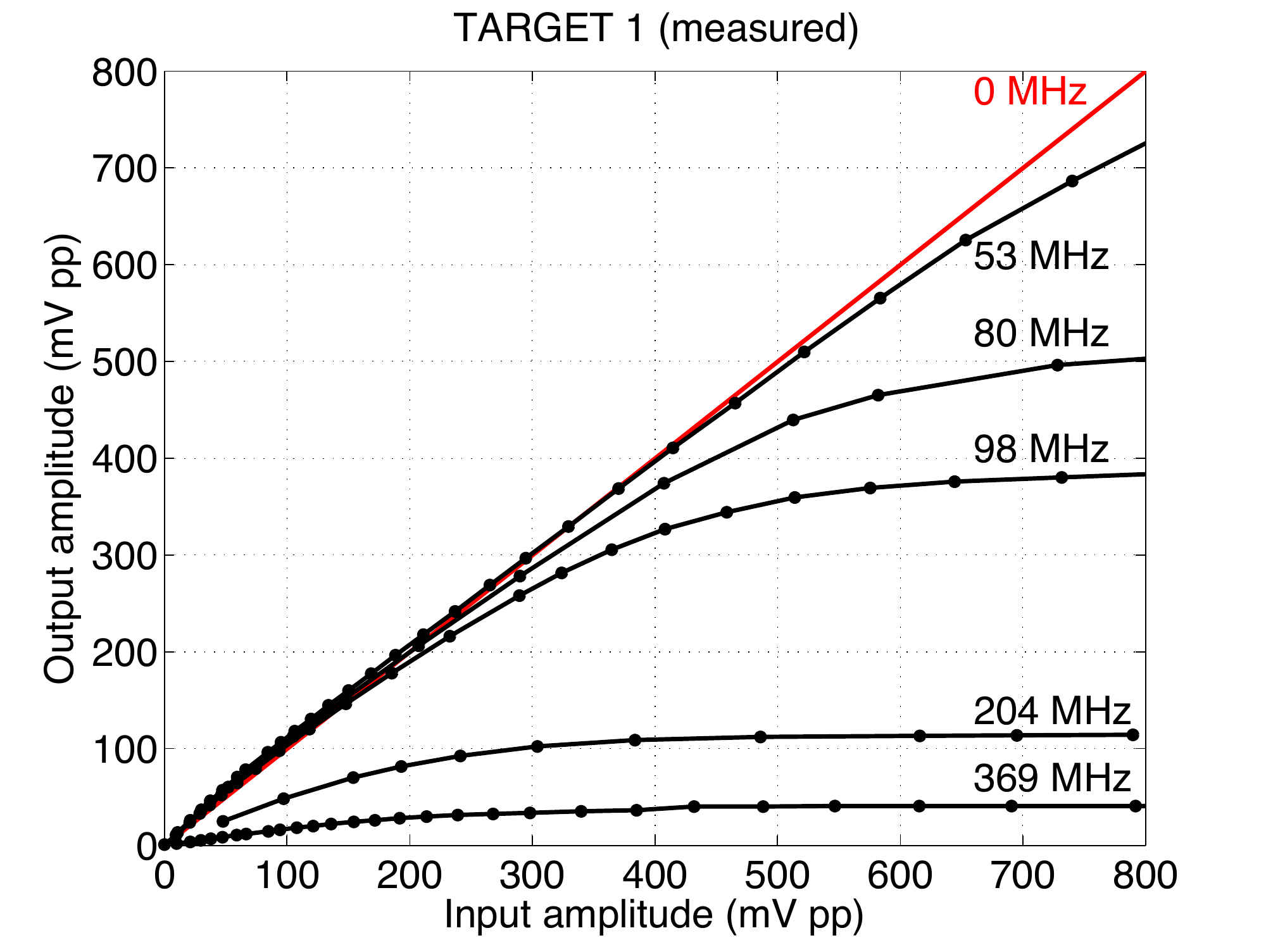}
}
\subfigure[]{
\label{target1simulation}
\centering
% NOTE I am using the simulation results *with* bond wire effects included, not the results without them.
\noindent\includegraphics[width=0.48\textwidth]{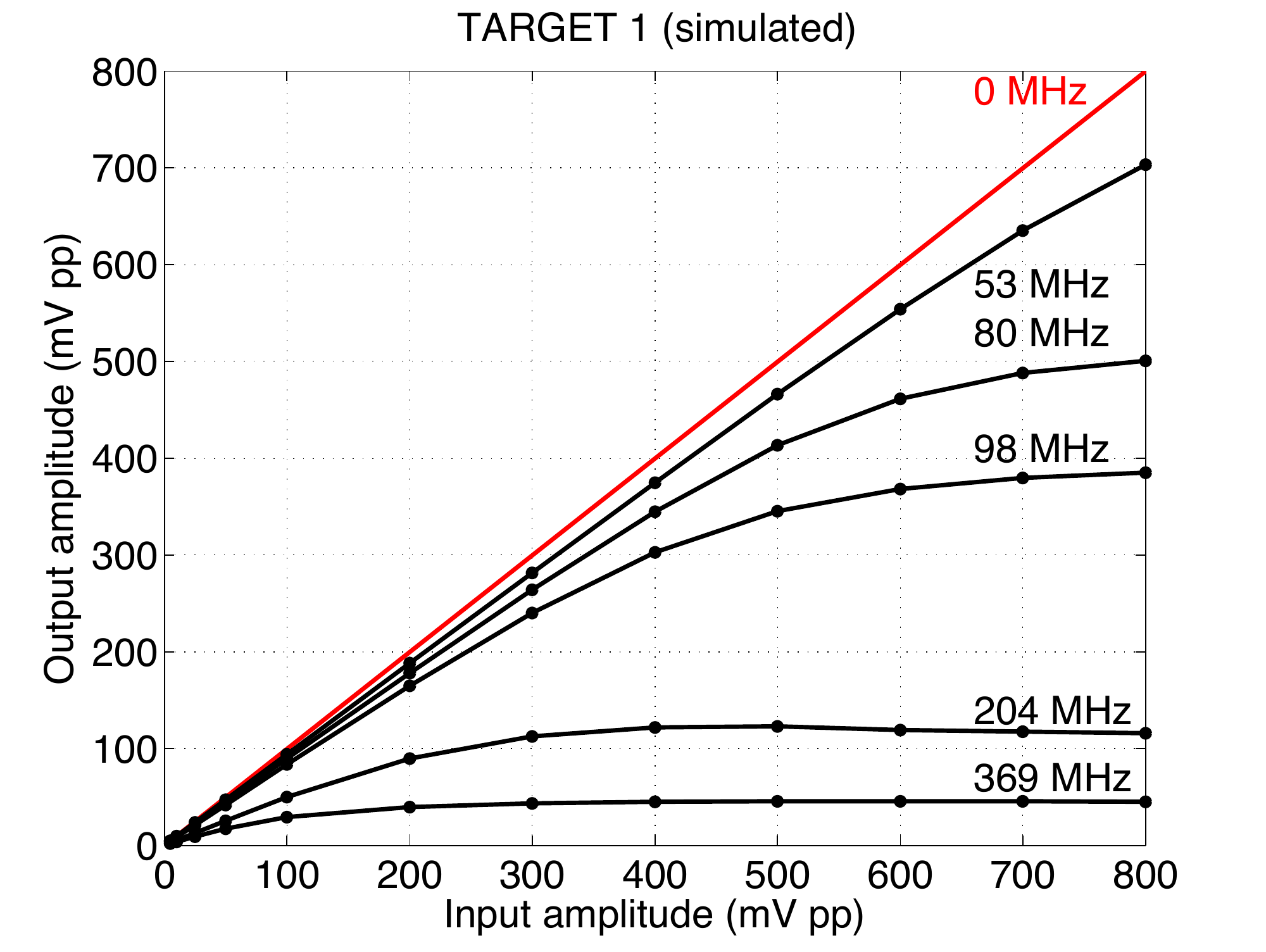}
}
\subfigure[]{
\label{target2saturation}
\centering
\noindent\includegraphics[width=0.48\textwidth]{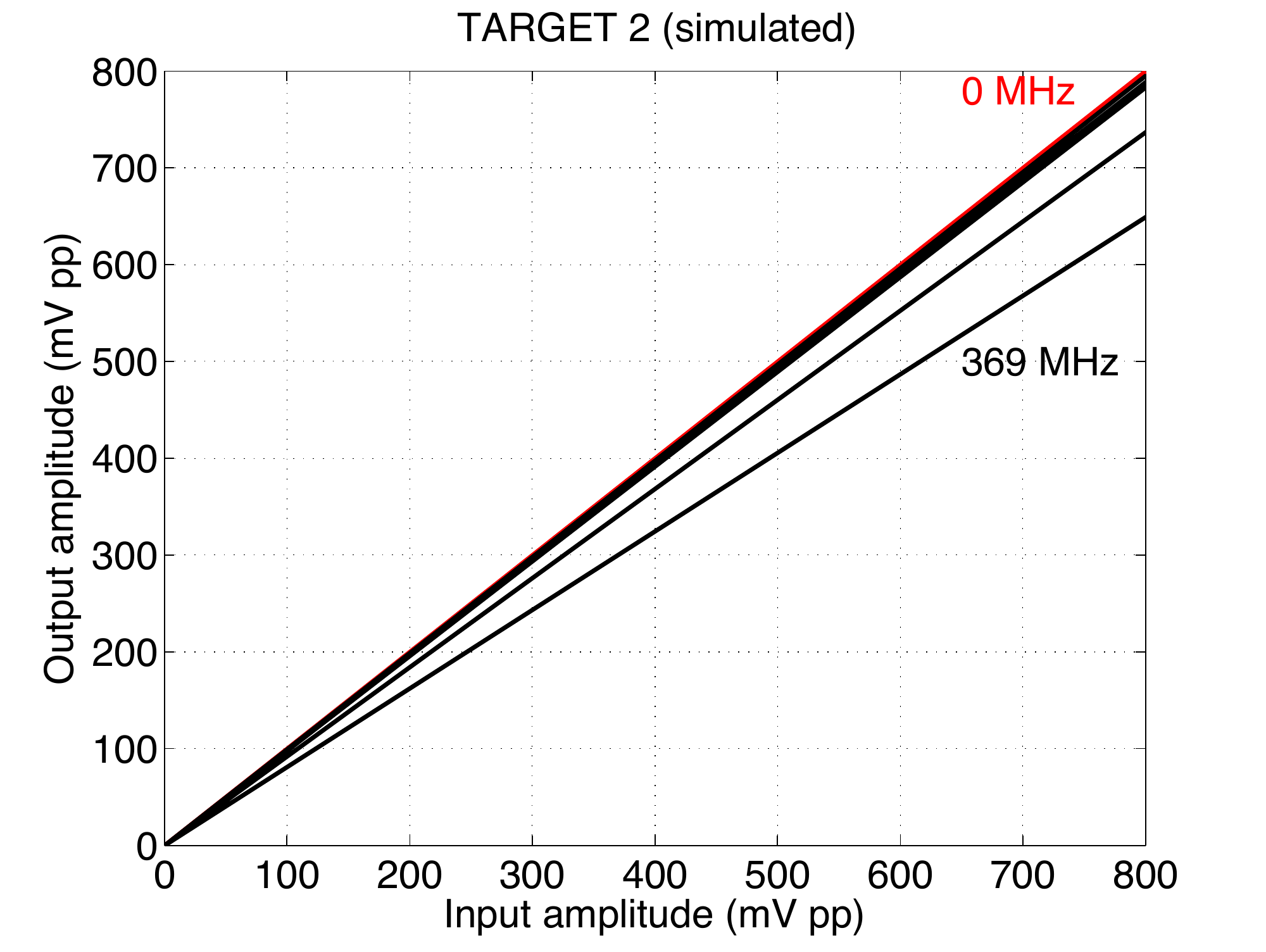}
}
\caption{Saturation of the TARGET~1 transfer function for sinusoidal signals at high frequency and large amplitude, as measured \subref{overlaySaturation} and simulated \subref{target1simulation}.  For low amplitude signals, the slope of the curves is independent of amplitude but dependent on frequency: this is the finite bandwidth of the chip.  Additionally, the curves roll over at high amplitude and do so most at high frequency: this is the saturation effect.  Measurements and simulations were made with sinusoids.  The DC limit (0 MHz) is shown for reference.  The saturation has been corrected in the design of TARGET~2.  Good agreement between the measured and simulated response of TARGET~1 provides confidence in our simulation of the TARGET~2 response \subref{target2saturation}.  In TARGET~2 the bandwidth effect is reduced and the saturation effect is eliminated.}
\label{saturation}
\end{center}
\end{figure*}

\subsection{AC saturation}

Section~\ref{transferFunction} presented the transfer function measured for DC voltages.  An ideal digitizer would feature the same transfer function for arbitrarily high signal frequencies.  In reality, the input buffer amplifiers cannot slew fast enough to keep up with the input signal if the signal frequency and amplitude are large.

We quantified this AC saturation effect by measuring the transfer function at several different input frequencies, as shown in Figure~\ref{saturation}.  The measurement was made by generating a sinusoid with a voltage-controlled oscillator and splitting the signal with a balanced splitter.  One output was routed to an oscilloscope for monitoring and the other was connected to the TARGET~1 chip.  The capacitor-dependent DC transfer functions were used to convert the recorded ADC counts to voltages, and a sinusoid was fit to each recorded waveform to determine the sinusoid amplitude as measured by the TARGET~1 chip.  The recorded amplitude can then be compared to the input amplitude (calibrated with the oscilloscope) as a function of both input amplitude and signal frequency, as shown in Figure~\ref{saturation}.

At low frequency the DC transfer function is recovered, but saturation is apparent for high-frequency, high-amplitude signals.  In addition to this saturation behavior, at high frequency the TARGET~1 gain is lower than at low frequency, even for small-amplitude signals.  This is visible in Figure~\ref{saturation} as a decrease in the slope of the curves, at low amplitude, as the frequency increases.  This is due to the finite analog bandwidth of the input buffer amplifiers, and is quantified in Section~\ref{bandwidthSection}.

The AC saturation effect has been addressed in the design of TARGET~2 and simulations indicate it is removed (see Figure~\ref{saturation}).

\begin{figure*}[]
\begin{center}
\subfigure[]{
\label{target1}
\centering
\noindent\includegraphics[width=0.48\textwidth]{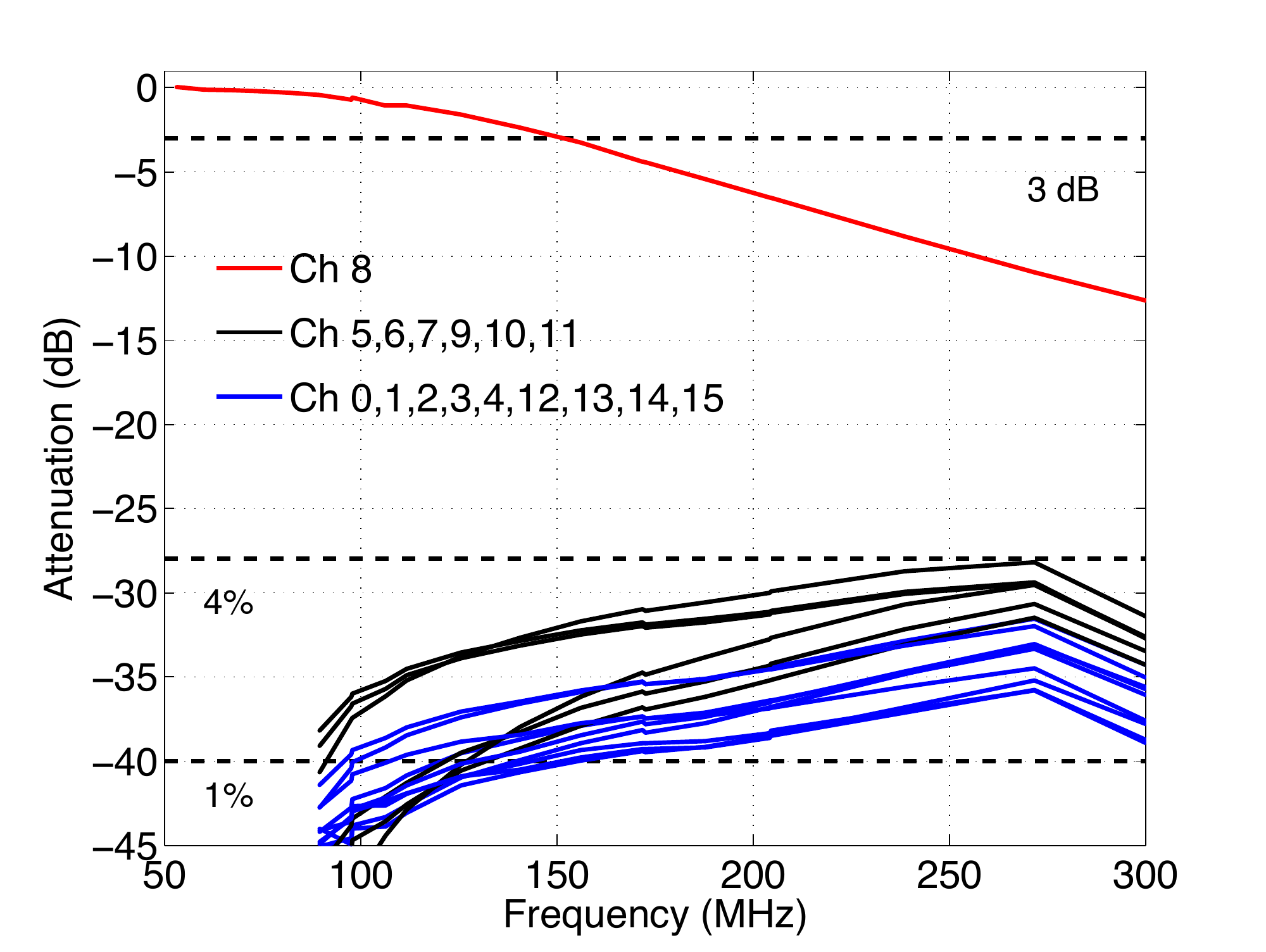}
}
\subfigure[]{
\label{target2}
\centering
\noindent\includegraphics[width=0.48\textwidth]{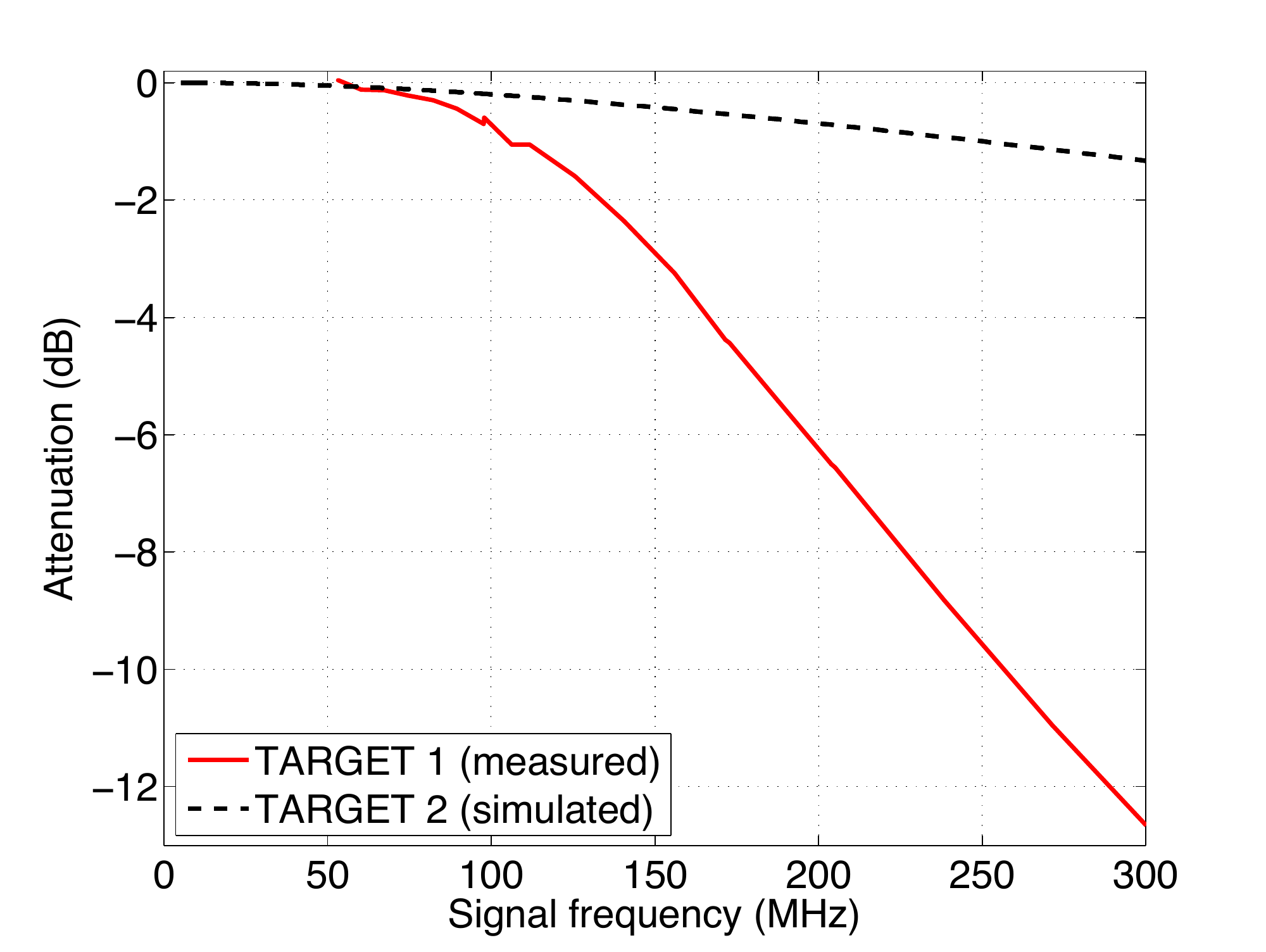}
}
\caption{\subref{target1} TARGET~1 analog bandwidth and cross talk.  A sinusoidal signal was input to Channel 8 of the evaluation board, and all 16 channels were digitized.  A balanced splitter was used to calibrate the signal amplitude on an oscilloscope simultaneously with the TARGET~1 recordings.  The ratio between the digitized and oscilloscope amplitudes is plotted for each channel as a function of frequency.  To measure the attenuation on Channel 8 (the signal channel), a small (100~mVpp) input amplitude was used in order to avoid saturation effects.  For the other channels (the cross talk channels), a large (750~mVpp) input amplitude was used in order to detect the cross talk above noise.  For all runs the internal termination of the TARGET~1 chip was set to 10~k$\Omega$.  The cross talk is at the 1-4\% level, and the 3~dB analog bandwidth is 150~MHz.  As expected, cross talk decreases for channels that are located physically farther from the signal channel: Channel 8 is plotted in red, channels at intermediate distances are plotted in black, and channels at the greatest distances are plotted in blue.  \subref{target2} Measured TARGET~1 frequency response compared with simulated TARGET~2 response, showing that TARGET~2 is designed to have a significantly larger ($>$~380~MHz) bandwidth.}
\label{plotCrossTalk}
\end{center}
\end{figure*}

\subsection{Analog bandwidth and cross talk}
\label{bandwidthSection}
A sinusoidal signal, generated by a voltage-controlled oscillator (VCO), was connected to Channel 8 and all 16 channels were recorded.  The VCO signal frequency was varied and several VCOs were used to span a wide range of frequencies. Figure~\ref{plotCrossTalk} shows the bandwidth and cross talk measured with this method.  The 3~dB bandwidth is 150~MHz.  A small amount of cross talk (between 1\% and 4\%, varying with frequency and channel) is observed, largest in the channels with traces closest to the signal channel on the evaluation board.  The nominal cross talk value is determined by comparing the not-connected channels with the signal channel at the 3 dB frequency (i.e. after attenuation of the signal channel by the finite bandwidth) and yields a value $\leq4$\% for all channels.

\begin{figure}[]
\centering
\noindent\includegraphics[width=0.48\textwidth]{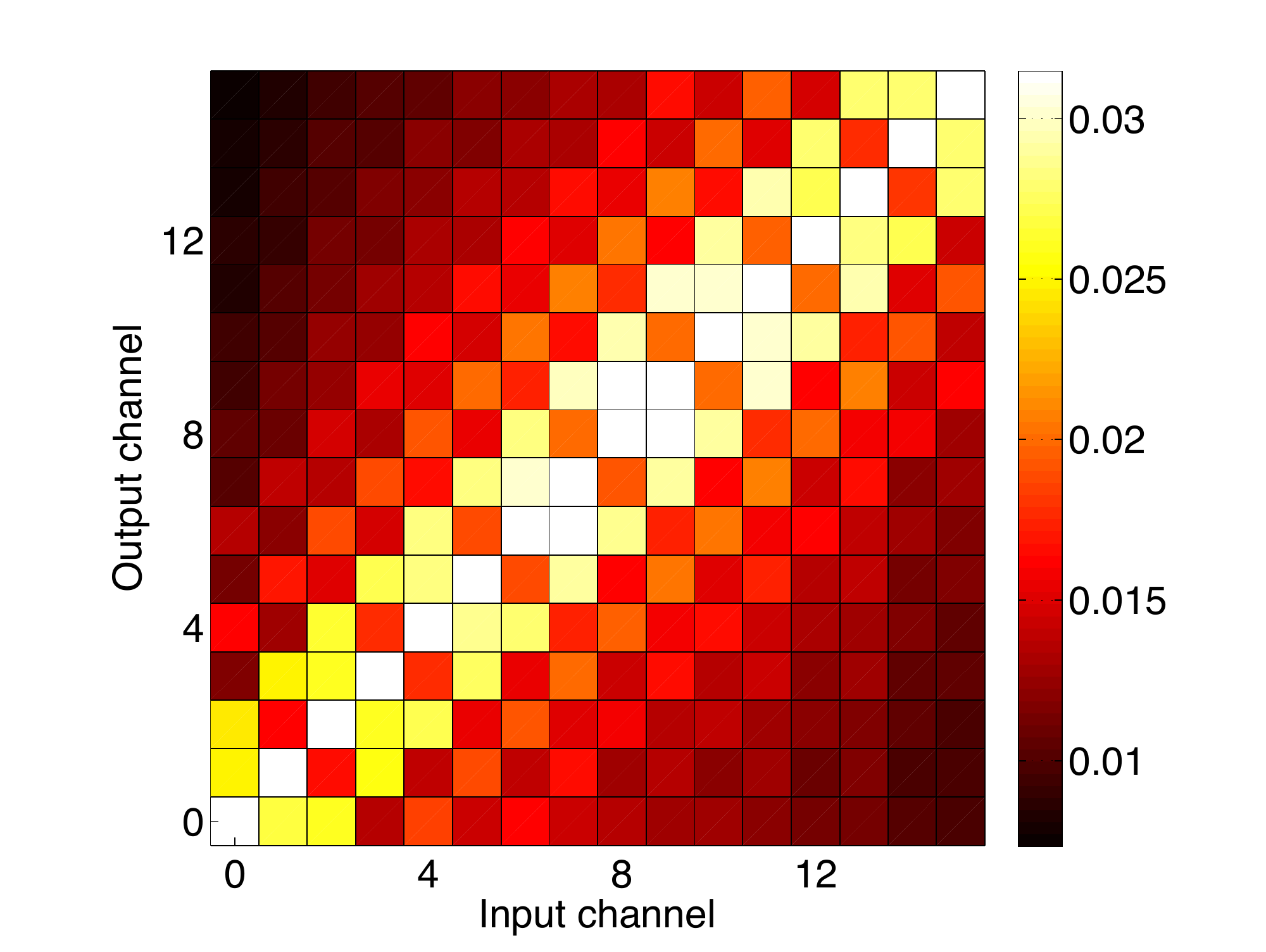}
% Figure generated with crossTalkTable.m
\caption{Cross-talk matrix at 156~MHz.  The cross talk is defined as the ratio between output amplitude and input amplitude, not as the ratio between output amplitude and output amplitude on the signal channel.  Diagonal entries are off-scale and have been set to full-scale.  The matrix is symmetric.  Cross talk values are in the 1\% to 3\% range.  The cross talk level depends strongly on the distance between input and output channels.  It depends weakly on whether the two are in the same even/odd group or not, which induces a checkerboard pattern in the matrix.}
\label{crossTalkMatrix}
\end{figure}

In addition to the cross talk measurement shown in Figure~\ref{plotCrossTalk} for a range of frequencies and a single input channel, we measured the full cross-talk matrix for all input/output channel combinations at 156~MHz (near the 3~dB frequency).  The cross talk trend indicated in Figure~\ref{plotCrossTalk} for Channel 8 input continues for other channels: the cross talk decreases as the difference between channel numbers increases.  Because Channel 8 is in the middle of the channels, it therefore corresponds to the worst case of cross talk.  Signals input to channels at one end of the range or the other induce a small amount of cross talk on channels at the opposite end.  The full cross-talk matrix at 156~MHz is shown in Figure~\ref{crossTalkMatrix}.

\subsection{Noise}

The noise level was measured by taking high-statistics recordings of the pedestal voltage (set to 1.3~V for this test) with each capacitor.  The onboard pedestal voltage was used, rather than supplying an external DC voltage, in order to minimize external noise injection.  The standard deviation of recorded amplitudes provides an estimate of the noise level.  This noise level was found to be capacitor-dependent, with a mean value of 3.3 ADC counts (2.2~mV).  This includes common-mode noise injected by the onboard voltage source.  Removing this common-mode noise statistically, we estimate the intrinsic noise level of the TARGET~1 chip to be 2.6 counts or 1.4 bits. This measurement was made with the same ramp speed used to measure the transfer function reported in Section~\ref{transferFunction}.

%This is for the configuration XXX.

%TODO measure noise level vs. voltage (for onboard Vped, not for injected external voltage)

Thermal noise in the capacitors of the TARGET~1 storage buffer contributes significantly to the overall noise level in signal digitization. Given a storage cell capacitance of 42.0~fF, the thermal noise level at an operating temperature of 300 K is approximately 0.31 mV, using the relation

\begin{equation}
V_\mathrm{RMS}=\sqrt{k_\mathrm{B}T/C},
\end{equation}

\noindent where $k_\mathrm{B}$ is Boltzmann's constant, $T$ is the operating temperature, and $C$ is the storage cell capacitance.

\subsection{Dynamic range}

The transfer function and noise results, measured using the same ramp speed configuration described above, can be combined to determine the effective (noise-subtracted) dynamic range of the digitizer.  The operating range in this configuration is 0.6 to 1.6~V (939.2 to 2399.7 ADC counts).  This corresponds to a dynamic range of 1460.5 ADC counts, or 10.5 bits.  The mean (averaging over all capacitors) noise level is 2.6 ADC counts after common-mode noise subtraction, corresponding to 1.4 bits of noise.  The effective dynamic range is therefore $10.5-1.4=9.1$~bits.  This can be increased (or decreased) by changing the ramp speed, resulting in a correspondingly larger (or smaller) dead time.

%%%%%%%%%%%%%%%%%%%%%% Future Directions %%%%%%%%%%%%%%%%%%%%%%%%%%%%%%%%%%%%

%\begin{figure}[ht]
\begin{figure*}[]
\vspace*{0mm}
%\noindent\includegraphics[width=0.48\textwidth]{TARGET_evall_pcb}
%\centerline{\psfig{file=TARGET_evall_pcb.eps,width=3.2in}}
\centering
\noindent\includegraphics[width=0.95\textwidth, angle=0]{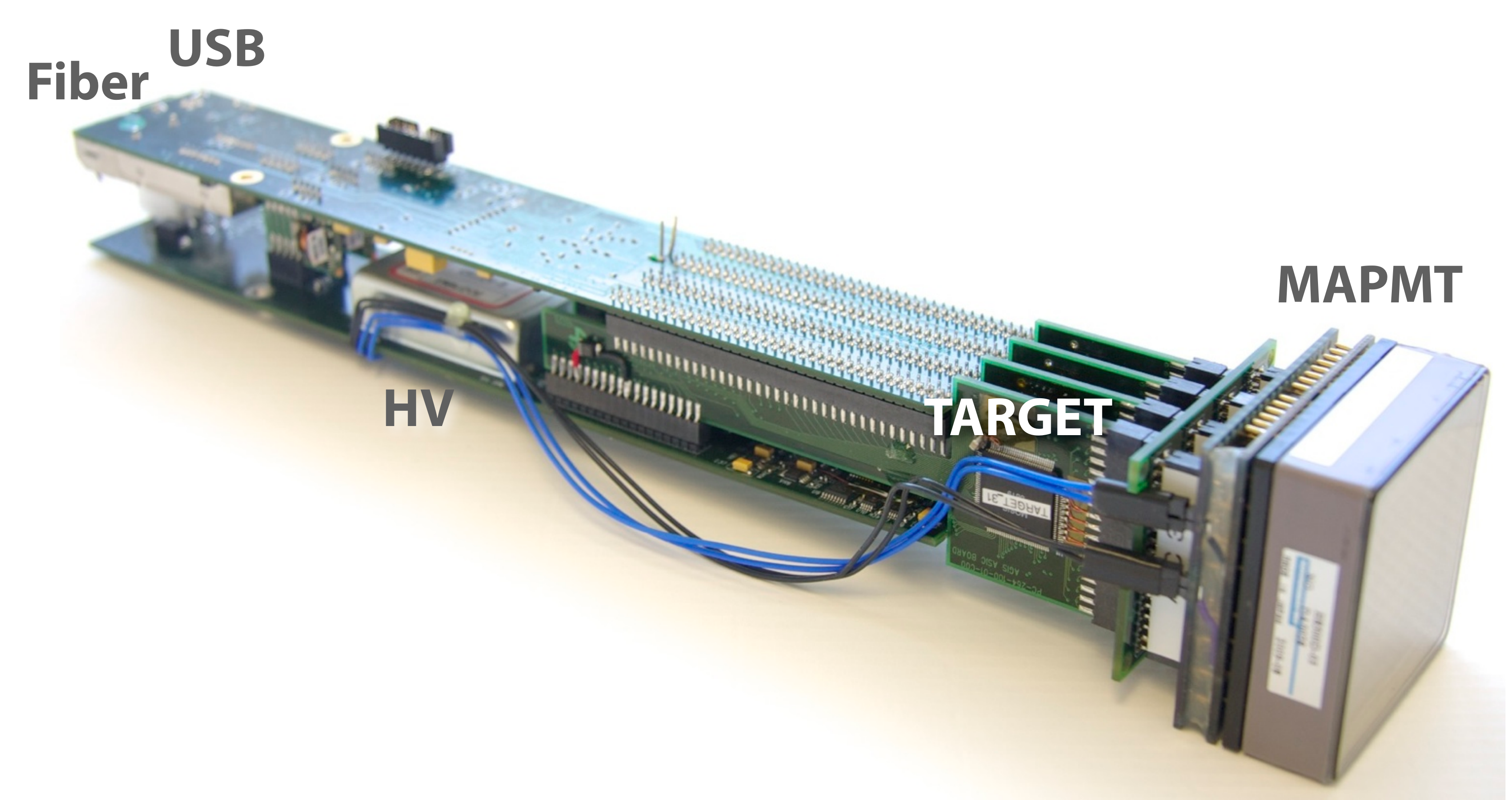}
\vspace*{0mm}
\caption{Prototype camera module.  On the right is a MAPMT (Hamamatsu model H8500D-03, eight pixels by eight pixels).  This connects to four TARGET~1 daughter boards (vertical).  In addition to the four TARGET~1 boards there are two main boards (horizontal) which together have an FPGA, a high voltage supply, a USB interface, and a fiber optic interface.  The fiber provides the primary interface and the USB is used for initial testing.}
\label{cameraModule}
\end{figure*}

\section{Camera module prototype}
\label{sec:camera_module}

In addition to the evaluation board, which features one FPGA and one TARGET~1 chip, a prototype front-end camera module board has been designed and produced.  A photograph of this module is shown in Figure~\ref{cameraModule}.  The module features four TARGET~1 chips and one FPGA (Xilinx Virtex XC5VLX30T), along with a high-voltage supply, a USB interface, and a fiber optic interface.  Similarly to the evaluation board, DAQ software on a PC communicates with the FPGA firmware, and the FPGA controls the four TARGET~1 chips.

While a USB interface was included for initial testing, the primary interface is a fiber optic that carries both trigger signals and data at a high rate from the camera module to a backplane.  A sustained readout rate of 3.3~kHz has been achieved (operating the fiber interface at 1 Gbps) from the camera module, reading out all 64 photosensor channels of the module, with 48 samples per channel, for each event.  With a single-event size of 6556 bytes, this corresponds to a data rate of 21.6~MB/sec.  The FPGA firmware can be upgraded to achieve 2 Gbps and a 6.6~kHz event rate.  A serial protocol is used to carry trigger signals and data on the same fiber.

With four TARGET~1 chips and a single FPGA, the camera module board can read out 64 photosensor channels with a single compact module.  The module is well suited to interface with a single 64-channel MAPMT.  It could also read out 64 silicon photomultiplier (multi-channel photon counter) pixels or individual PMTs.  Dozens of such modules can be used to provide the front-end readout electronics for a telescope camera.  They can each connect via fiber optic to a backplane that performs telescope-level trigger logic and collects data to transfer from the telescope.

%Reading out three blocks (48 samples) per trigger has been demonstrated with the camera module board.  This may be sufficient to capture the full photomultiplier waveform while requiring less readout time and transfer time downstream than four blocks (64 samples) per trigger as is done with the evaluation board.

\section{TARGET~2}
\label{sec:TARGET2}

The basic performance of the first TARGET ASIC prototype (TARGET~1) for application in VHE gamma-ray astronomy has been demonstrated.  Design and fabrication of a second generation of the TARGET chip (TARGET~2) is complete and its performance will be similarly characterized.  The specifications for TARGET~2 are shown in Table ~\ref{specTable}.  TARGET~2 features 16,384 sample capacitors per channel instead of 4,096 (in order to increase the buffer depth from 4.1~$\mu$s to 16.4~$\mu$s at 1 GSa/s) as well as 512 Wilkinson ADCs instead of 32, which allows more parallel digitization in order to decrease the dead time.  The increased buffer depth will provide sufficient time to perform sophisticated multi-telescope trigger logic within the IACT array, allowing improvements in hadronic shower background rejection and enabling a lower energy threshold.

Waveform sampling and storage are decoupled in TARGET~2 by using a two-stage transfer system: a small sampling buffer transfers to a large analog ring buffer for storage.  Sampling is performed by a small ping-pong buffer consisting of two lanes of 32 samples each.  One lane samples while the other transfers to a large storage buffer (the second stage).  The storage buffer features 8 rows with 64 columns per row and one block of 32 samples per (row, column) combination, for a total of 16,384 storage cells.  The two-stage system increases the analog bandwidth, lowers digitization noise, and reduces the number of timing calibration constants required.

Noise and cross talk are reduced by using pseudo-differential sampling preamps.  The control signals to start and clear the Wilkinson ramp signals are provided by the FPGA in TARGET~1 but are incorporated into the ASIC itself in TARGET~2; these allow configuration of a delay between the counter start and ramp start.  The ramp start voltage and speed are controlled by external control voltages for both chips.  Self-triggering capabilities have also been expanded in TARGET~2: while the TARGET~1 trigger is the OR of the 16 individual channel comparators, TARGET~2 provides 4 separate trigger signals each of which is the analog sum of 4 channels, as well as a fifth trigger signal which is the analog sum of all 16 channels.  This enables more sophisticated, fine-grained trigger logic.  Finally, TARGET~2 includes internal threshold DACs, transimpedance amplifiers, and a DAC to control the width of the trigger signal (accomplished with the external WBIAS control voltage in TARGET~1).

%%%%%%%%%%%%%%%%%%%%%% Summary %%%%%%%%%%%%%%%%%%%%%%%%%%%%%%%%%%%%

\section{Conclusion}
\label{sec:summary}

We have described the architecture of an ASIC tailored to the low-cost, high-multiplicity requirements to read out the photodetectors of the Cherenkov Telescope Array.  The 16-channel TARGET~1 chip has been evaluated and meets nearly all requirements for such an application.  In addition, we produced a front-end camera module prototype to demonstrate integration of TARGET into a system capable of reading out an IACT focal plane.  The camera module can currently read out events at 3.3~kHz through a fiber optic link, upgradable in firmware to 6.6~kHz.

A second-generation digitizing ASIC, TARGET~2, has been designed and fabricated and will be characterized in the same way as reported here for TARGET~1.  TARGET~2 incorporates a deeper buffer architecture, reduced dead time, more functionality integrated into the ASIC itself rather than the companion FPGA, and expanded triggering capabilities which will enable sophisticated trigger logic at both the single-telescope and multi-telescope level.  TARGET~2 is expected to meet the requirements for the IACT application.

The deep buffer and high channel multiplicity of the TARGET design enable flexible triggering and low per-channel cost, power consumption, and weight.  In addition to CTA, the TARGET ASIC's programmable input termination, narrow digitization selection window, fast signal conversion, and large channel multiplicity make it useful for a number of other applications.  These include drift chamber readout, collider detectors for kaons and muons, and a variety of focal plane array readout systems.

%%%%%%%%%%%%%%%%%%%%%% Summary %%%%%%%%%%%%%%%%%%%%%%%%%%%%%%%%%%%%

\section{Acknowledgments}
We are grateful for essential contributions to the design from J.~Buckley (Washington University) and the CTA-US camera working group.  We also acknowledge excellent system engineering by L.~Sapozhnikov (SLAC National Accelerator Laboratory) and help in the laboratory from R.~Buehler (KIPAC) and valuable discussions with D.~Tosi (Stanford University).  Testing was supported in part by Department of Energy Advanced Detector Research Award \# DE-FG02-06ER41424. This work is supported by the Department of Energy, Laboratory Directed Research and Development funding, under contract \# DE-AC02-76SF00515.  J.~V. is supported by a Kavli Fellowship from the Kavli Foundation.  A.~O. is supported by Grant-in-Aid for JSPS Fellows.  A.~S. is supported by the STAR (STEM Teacher and Reacher) program.  S.~F.~ acknowledges generous support by the KIPAC Enterprise funds and by the Mel Schwarz Award from Stanford University.  K.~B.~ is supported by a Stanford Graduate Fellowship.  We are grateful for valuable suggestions by two anonymous referees.

%%%%%%%%%%%%%%%%%%%%%%%%%%% Bibliography %%%%%%%%%%%%%%%%%%%%%%%%%%%%%%%%%

\bibliography{target1}
\bibliographystyle{elsart-num}

%%%%%%%%%%%%%%%%%%%%%%%%%%%%%%%%%%%%%%%%%%%%%%%%%%%%%%%%%%%%%%%%%%%%%%%%
\end{document}